\documentclass[times,10pt,twocolumn]{article}  

\usepackage{latex8} 
\usepackage{times} 
\usepackage{epsfig} 
\usepackage{amssymb}  
\usepackage{latexsym}  
  
\usepackage[PostScript=dvips]{diagrams}  
\input {diagrams.sty}
\newarrow{Toto}<--->
\newarrow{Is}===== 
\newarrow{Dotsto}....>
\newarrow{Dots}.....

\begin{document}  

\newcommand{\bpf}{\noindent {\bf Proof:} }
\def\endproof{\hfill$\Box$} 

\newcommand{\cnot}{\textsc{\footnotesize CNOT}}
\newcommand{\bit}{\begin{itemize}}
\newcommand{\eit}{\end{itemize}\par\noindent}
\newcommand{\ben}{\begin{enumerate}}
\newcommand{\een}{\end{enumerate}\par\noindent}
\newcommand{\beq}{\begin{equation}}
\newcommand{\eeq}{\end{equation}\par\noindent}
\newcommand{\beqa}{\begin{eqnarray*}}
\newcommand{\eeqa}{\end{eqnarray*}\par\noindent} 
\newcommand{\beqn}{\begin{eqnarray}}  
\newcommand{\eeqn}{\end{eqnarray}\par\noindent}           

\newcommand{\dd}{\llcorner}
\newcommand{\sdot}{\bullet}
\newcommand{\ddd}{\lrcorner}
\newcommand{\uu}{\ulcorner} 
\newcommand{\uuu}{\urcorner}
\newcommand{\ot}{\otimes}

\newcommand{\HH}{\mathcal{H}}
\newcommand{\KK}{\mathcal{K}}
\newcommand{\ie}{\textit{i.e.}\ }
\newcommand{\Zero}{\mathbf{0}}
\newcommand{\CC}{\mathbf{C}}
\newcommand{\II}{{\rm I}}
\newcommand{\PP}{{\rm P}}
\newcommand{\bra}{\mathsf{bra}}
\newcommand{\FdHilb}{\mathbf{FdHilb}}

\newcommand{\neoiota}{\psi}

\newtheorem{Th}{Theorem}[section] 
\newtheorem{theorem}[Th]{Theorem} 
\newtheorem{proposition}[Th]{Proposition} 
\newtheorem{lemma}[Th]{Lemma} 
\newtheorem{corollary}[Th]{Corollary} 
\newtheorem{definition}[Th]{Definition} 
\newtheorem{example}[Th]{Example} 

\title{A categorical semantics of quantum protocols}  

\author{Samson Abramsky and Bob Coecke\\ \\
Oxford University Computing Laboratory,\\ 
Wolfson Building, Parks  
Road, Oxford OX1 3QD, UK.\\ {\sf\small samson.abramsky $\cdot$
bob.coecke@comlab.ox.ac.uk}\\
}  

\maketitle 
\thispagestyle{empty} 

\begin{abstract}
We study quantum information and computation from a novel point of
view.  Our approach is based on recasting the standard axiomatic
presentation of quantum mechanics, due to von Neumann {\rm\cite{vN}}, at a more
abstract level, of compact closed categories with biproducts.  We
show how the essential structures found in key quantum information
protocols such as teleportation {\rm\cite{BBC}}, logic-gate teleportation
{\rm\cite{Gottesman}}, and entanglement swapping {\rm\cite{Swap}} can be captured
at this abstract level.  Moreover, from the combination  of the --- apparently
purely qualitative --- structures of compact closure and biproducts there
emerge `scalars' and a  `Born rule'.  This abstract and structural
point of view opens up new possibilities for describing and
reasoning about quantum systems. It also shows the degrees of
axiomatic freedom: we can show what requirements are placed on the
(semi)ring of scalars ${\bf C}(\II,\II)$, where ${\bf C}$ is the category and
$\II$ is the tensor unit, in order to perform various protocols such as 
teleportation. Our formalism captures both the information-flow
aspect of the protocols {\rm\cite{Coe1,Coe2}}, and the branching
due to quantum indeterminism. This contrasts with the standard
accounts, in which the classical information flows are `outside'
the usual quantum-mechanical formalism.  We give detailed formal descriptions and proofs of correctness of the example protocols.
\end{abstract}
 
\Section{Introduction}   

Quantum information and computation is concerned with the use of
quantum-mechanical systems to carry out computational and
information-processing tasks \cite{Nielsen}.
In the few years that this approach has been studied, a number of 
remarkable concepts  and results have emerged.
Our particular focus in this paper is on \emph{quantum information
  protocols}, which exploit quantum-mechanical effects in an essential 
way. The particular examples we shall use to illustrate our approach
will be \emph{teleportation} \cite{BBC}, \emph{logic-gate teleportation}
\cite{Gottesman}, and \emph{entanglement swapping} \cite{Swap}. The ideas illustrated in
these protocols form the basis for novel and potentially very
important applications to secure and fault-tolerant communication and
computation \cite{BEZ,Gottesman,Nielsen}.
 
We now give a thumbnail sketch of teleportation to motivate our
introductory discussion. (A more formal `standard' presentation is
given in Section~2. The --- radically different --- presentation in our new
approach appears in Section~9.)
Teleportation involves using an entangled pair of qubits $(q_A , q_B
)$ as a kind of communication channel to transmit an unknown qubit $q$ 
from a source $A$ (`Alice') to a remote target $B$ (`Bob'). $A$ has $q$
and $q_A$, while $B$ has $q_B$. We firstly entangle $q_A$ and $q$ at
$A$ (by performing a suitable unitary operation on them), and then
perform a measurement on $q_A$ and $q$.\footnote{This measurement can be performed in the
standard `computational basis'. The combination of unitary and
measurement is equivalent to measurement in the `Bell basis'.} This forces a
`collapse' in $q_B$ because of its entanglement with $q_A$. We then
send two classical bits of information from $A$ to $B$, which encode
the four possible results of the measurement we performed on $q$ and
$q_A$. Based on this classical communication, $B$ then performs 
a `correction' by applying one of four possible operations (unitary
transformations) to $q_B$, after which \emph{$q_B$ has the same state
  that $q$ had originally}. (Because of the measurement, $q$ no longer
has this state --- the information in the source has been `destroyed' in 
transferring it to the target). It should be born in mind that the
information required to specify $q$ is an arbitrary pair of complex
numbers $(\alpha ,\beta)$ satisfying $|\alpha|^2 + |\beta|^2 = 1$, so achieving this information transfer with just two
classical bits is no mean feat!

Teleportation is simply the most basic of a family of quantum
protocols, and already illustrates the basic ideas, in particular the
use of \emph{preparations of entangled states} as carriers for information
flow, performing \emph{measurements} to propagate information, using
\emph{classical information} to control branching behaviour to ensure the
required behaviour despite quantum indeterminacy, and performing local 
data transformations using \emph{unitary operations}. (Local here means
that we apply these operations only at $A$ or at $B$, which are
assumed to be spatially separated, and not simultaneously at both).

Our approach is based on recasting the standard axiomatic presentation 
of Quantum Mechanics, due to von Neumann \cite{vN}, at a more abstract
level, of \emph{compact closed categories with biproducts}.
Remarkably enough, all the essential features of quantum protocols
mentioned above find natural counterparts at this abstract level ---
of which the standard von Neumann presentation in terms of Hilbert
spaces is but one example. More specifically:
\begin{itemize} 
\item The basic structure of a symmetric monoidal category allows
  \emph{compound systems} to be described in a resource-sensitive
  fashion (cf.~the `no
  cloning' and `no deleting' theorems of quantum mechanics \cite{Nielsen}).
\item The compact closed structure allows \emph{preparations and 
    measurements of entangled states} to be described, and their
key
  properties to be proved.
\item Biproducts allow \emph{indeterministic branching, classical
communication  and superpositions} to
  be captured.
\end{itemize}
We are then able to use this abstract
setting to give precise formulations of teleportation, logic gate
teleportation, and entanglement swapping, and to prove correctness of
these protocols --- for example, proving correctness of teleportation
means showing that the final value of $q_B$ equals the initial value
of $q$. Moreover, from the combination  of the---apparently purely qualitative---structures
of compact closure and biproducts  there emerge \em scalars \em and a  \em
Born rule\em.

One of our main concerns is to replace ad hoc calculations with bras and kets,
normalizing constants, unitary matrices etc. by conceptual
  definitions and proofs. This allows general underlying structures to 
  be identified, and general lemmas to be proved which encapsulate key formal
  properties. The compact-closed level of our axiomatization allows
  the key \emph{information-flow properties} of entangled systems to be
  expressed. Here we are directly abstracting from the more concrete
  analysis carried out by one of the authors in \cite{Coe1,Coe2}. The advantage
of our abstraction is 
  shown by the fact that the extensive linear-algebraic calculations
  in \cite{Coe1} are replaced by a few simple conceptual lemmas, valid in an
  arbitrary compact closed category. We are also able to reuse the
  template of definition and proof of correctness for the basic
  teleportation protocol in deriving and verifying logic-gate
  teleportation and entanglement swapping.

The compact-closed level of the axiomatization allows information flow 
along any branch of a quantum protocol execution to be described, but
it does not capture the \emph{branching} due to measurements and quantum
indeterminism. The biproduct structure allows this branching behaviour 
to be captured. Since biproducts induce a (semi)additive structure,
the superpositions characteristic of quantum phenomena can be
captured at this abstract level. Moreover, the biproduct structure
interacts with the compact-closed structure in a non-trivial
fashion. In particular, the \emph{distributivity} of tensor product
over biproduct allows classical communication, and the dependence of
actions on the results of previous measurements (exemplified
in teleportation by the dependence of the unitary correction on the
result of the measurement of $q$ and $q_A$), to be captured within
the formalism. In this respect, our formalism is \emph{more
comprehensive} than the standard von Neumann axiomatization. In the
standard approach, the use of measurement results to determine subsequent
actions is left informal and implicit, and hence not subject to
rigorous analysis and proof. As quantum protocols and  computations grow more
elaborate and complex, this point is likely to prove of increasing
importance. 

Another important point concerns the \emph{generality} of our
axiomatic approach. The standard von Neumann axiomatization fits
Quantum Mechanics perfectly, with no room to spare. Our basic setting
of compact closed categories with biproducts is general enough to
allow very different models such as $\mathbf{Rel}$, the category of
sets and relations. 
When we consider specific protocols such as teleportation, a kind of 
`Reverse Arithmetic' (by analogy with Reverse Mathematics \cite{RM})
arises. That is, we can characterize what requirements are placed on
the `semiring of scalars' ${\mathbf C} ({\rm I},{\rm I})$ (where ${\rm I}$ is the
tensor unit) in order for the protocol to be realized. This is often
much less than requiring that this be the field of complex numbers
(but in the specific cases we shall consider, the requirements are
sufficient to exclude $\mathbf{Rel}$).
Other degrees of axiomatic freedom also arise, although we shall not
pursue that topic in detail in the present paper.

The remainder of the paper is structured as follows. Section~2
contains a
rapid review of the standard axiomatic presentation of Quantum
Mechanics, and of the standard presentations of our example protocols.
Section~3 introduces compact closed categories, and presents the key
lemmas on which our analysis of the information-flow properties of
these protocols will be based. Section~4 relates this general analysis 
to the more concrete and specific presentation in \cite{Coe1}. Section~5
introduces biproducts.  Sections~6 and~7  present our abstract
treatments of  scalars and adjoints. Section~8 presents our abstract formulation of quantum mechanics.
Section~9 contains our
formal descriptions and verifications of the example protocols. 
Section~10 concludes.

\Section{Quantum mechanics and teleportation}    

In this paper, we shall only consider \emph{finitary} quantum
mechanics, in which all Hilbert spaces are finite-dimensional. This is 
standard in most current discussions of quantum computation and
information \cite{Nielsen},
and corresponds physically to considering only observables with finite
spectra, such as \emph{spin}. (We  refer briefly to the extension 
of our approach to the infinite-dimensional case in the
Conclusions.)    

Finitary quantum theory has the following basic ingredients (for
more details, consult standard texts such as \cite{Isham}).
\ben
\item[{\bf 1.}] The \em state space \em of the system is represented as a
  finite-dimensional Hilbert space $\HH$, \ie a finite-dimensional complex
  vector space with an inner product written $\langle \phi \mid \psi
  \rangle$, which is conjugate-linear in the first argument and linear 
  in the second.  A \emph{state} of a quantum system
corresponds to a one-dimensional subspace ${\cal A}$ of $\HH$, and is standardly
represented by a vector $\psi\in {\cal A}$ of unit norm. 
\item[{\bf 2.}] For informatic purposes, the basic type is that of
  \emph{qubits}, namely $2$-dimensional Hilbert space, equipped with a
  \em computational basis \em $\{|0\rangle,|1\rangle\}$.
\item[{\bf 3.}] \em Compound systems \em are described by tensor products of the
  component systems. It is here that the key phenomenon of
 \em entanglement \em arises, since the general form of a vector in $\HH_1
  \otimes \HH_2$ is\vspace{-1.20mm}
\[ 
\sum_{i=1}^n \alpha_i\cdot \phi_i \otimes \psi_i\vspace{-1.20mm} 
\] 
Such a vector may encode \emph{correlations} between the 
first and second components of the system, and cannot simply be
resolved into a pair of vectors in the component spaces.
\een
The \em adjoint \em to a linear map
$f:{\cal H}_1\to{\cal H}_2$ is the  linear map $f^\dagger:{\cal
H}_2\to{\cal H}_1$ such that, for all $\phi\in {\cal H}_2$ and  
$\psi\in {\cal H}_1$,
\[
\langle\phi \mid f(\psi)\rangle_{{\cal H}_2} = \langle f^\dagger(\phi) \mid
\psi\rangle_{{\cal H}_1}\,.
\]
\emph{Unitary
  transformations} are linear isomorphisms 
\[
U:{\cal H}_1\to{\cal H}_2
\] 
such that
\[
U^{-1}\!\!=U^\dagger:{\cal H}_2\to{\cal H}_1\,.
\]
Note that all such transformations 
\em preserve the inner product \em since,
for all $\phi,\psi\in{\cal H}_1$,
\[
\langle U(\phi) \mid U(\psi)\rangle_{{\cal H}_2}=\langle (U^\dagger U)(\phi) \mid
\psi\rangle_{{\cal H}_1}=\langle \phi \mid \psi\rangle_{{\cal H}_1}\,.
\]
\em Self-adjoint operators \em are linear transformations
\[
M :{\cal H}\to{\cal H}
\]
such that $M=M^\dagger$.

\ben
\item[{\bf 4.}] The \em basic data transformations \em are represented by unitary
transformations.  Note that all such data transformations are necessarily
\emph{reversible}.
\item[{\bf 5.}] The \em measurements \em which can be performed on the system are
represented by  self-adjoint operators.  
\een
The act of measurement itself consists of two parts:
\bit
\item[{\bf 5a.}] The observer is informed about the measurement outcome,
  which is a value $x_i$ in the spectrum $\sigma(M)$ of
the corresponding self-adjoint operator $M$. For convenience we assume
$\sigma(M)$ to be \em non-degenerate \em (linearly independent eigenvectors
have distinct eigenvalues). 
\item[{\bf 5b.}] The state of the system undergoes a change, represented by
  the action of the \em projector \em ${\rm P}_i$ arising from  the \em spectral decomposition \em 
\[ 
M=x_1\cdot {\rm P}_1+\ldots+x_n\cdot {\rm P}_n
\]
\eit
In this spectral decomposition the projectors ${\rm P}_i:{\cal H}\to{\cal
H}$ are  idempotent and self-adjoint, 
\[ 
{\rm P}_i\circ{\rm P}_i={\rm P}_i\quad {\rm and} \quad{\rm P}_i={\rm
P}_i^\dagger ,
\]
and
mutually orthogonal:
\[ {\rm P}_i\circ{\rm P}_j=0, \qquad  i\not=j . \]

This spectral decomposition always exists and is unique by the
\em spectral theorem \em for self-adjoint operators.  By our assumption
that $\sigma(M)$ was non-degenerate each projector ${\rm P}_i$ has a
one-dimensional subspace of
${\cal H}$ as its fixpoint set (which equals its image).

The probability of $x_i\in\sigma(M)$ being the actual outcome is given
by the \em Born rule \em which does not depend on
the value of $x_i$ but on ${\rm P}_i$ and the system state $\psi$, explicitly 
\[
{\sf Prob}({\rm
P}_i,\psi)=\langle\psi\mid {\rm P}_i(\psi)\rangle\,. 
\]
The
status of the Born rule within our abstract setting will emerge in
Section~8.
The derivable notions of \em mixed states \em and \em non-projective
measurements
\em will not play a significant r\^ole in this paper.

The values $x_1,\ldots,x_n$ are in effect merely labels distinguishing  
the projectors ${\rm P}_1,\ldots,{\rm P}_n$ in the
above sum. Hence we can abstract over them and think of
a measurement as a list of $n$ mutually orthogonal
projectors
$({\rm P}_1,\ldots,{\rm P}_n)$
where $n$ is the dimension of the Hilbert space. 

Although real-life experiments in many cases destroy the
system (e.g.~any measurement of a photon's location destroys it) measurements always have the same shape in
the quantum formalism.  When distinguishing between `measurements which preserve the system' and `measurements which
destroy the system' it would make sense to decompose a measurement explicitly in two components:
\bit
\item \em Observation \em consists of receiving the information on the outcome of the measurement, to be thought of
as specification of the index $i$ of the outcome-projector ${\rm P}_i$ in the above list. Measurements which destroy
the system can be seen as `observation only'$\!$. 
\item \em Preparation \em consists of producing the state ${\rm P}_i(\psi)$.
\eit
In our abstract setting these arise naturally as the two `building blocks' which are used to construct
projectors and measurements. 

We now discuss some important quantum protocols which we chose because 
of the key r\^ole entanglement plays in
them --- they involve both initially entangled states, and measurements against a basis of entangled states.

\subsection{Quantum teleportation}

The quantum teleportation protocol \cite{BBC} (see also \cite{Coe1} \S 2.3
and
\S 3.3) involves three qubits $a$, $b$ and $c$ (corresponding to $q$,
$q_A$ and $q_B$ respectively in our preliminary sketch in the Introduction). Qubit $a$ is in a state $|\phi\rangle$
and qubits $b$ and $c$ form an `EPR-pair', that is, their joint state is $|00\rangle+|11\rangle$. 
After spatial relocation (so that $a$ and $b$ are positioned at the
source $A$, while $c$ is positioned at the target $B$), one performs a \em Bell-base measurement
\em on $a$ and $b$, that is, a measurement such that each ${\rm P}_i$ projects on one of the
one-dimensional subspaces spanned by a vector in the \em
Bell basis\em: 
\[
\begin{array}{lcl}
b_1:={1\over \sqrt{2}}\cdot(|00\rangle\!+\!|11\rangle)&\ \ &
b_2:={1\over \sqrt{2}}\cdot(|01\rangle\!+\!|10\rangle)\vspace{2.5mm}\\ 
b_3:={1\over \sqrt{2}}\cdot(|00\rangle\!-\!|11\rangle)&\ \ &
b_4:={1\over \sqrt{2}}\cdot(|01\rangle\!-\!|10\rangle)\,.
\end{array} 
\]
This measurement can be of the type `observation only'. We observe the outcome of the measurement and depending on it
perform one of the unitary
transformations
\[
\begin{array}{lcl}
\beta_1:=\left(\begin{array}{rr} 
1&0\\
0&1
\end{array}\right)
&\quad&
\beta_2:=\left(\begin{array}{rr}
0&1\\
1&0
\end{array}\right)
\vspace{2.5mm}\\
\beta_3:=\left(\begin{array}{rr}
1&0\\
0&\!\!\!\!\!-\!1
\end{array}\right)
&\quad&
\beta_4:=\left(\begin{array}{rr}
0&\!\!\!\!\!-\!1\\
1&0
\end{array}\right) 
\end{array}\ 
\]
on $c$ --- $\beta_1,\beta_2,\beta_3$
are all self-inverse while
$\beta_4^{-1}=-\beta_4$. Physically, this requires transmission of two
classical bits, recording the outcome of
the measurement, from the location of
$a$ and $b$ to the location of $c$.

\vspace{1.2mm}\noindent{\footnotesize 
\hspace{-15pt}\begin{minipage}[b]{1\linewidth}  
\centering{\epsfig{figure=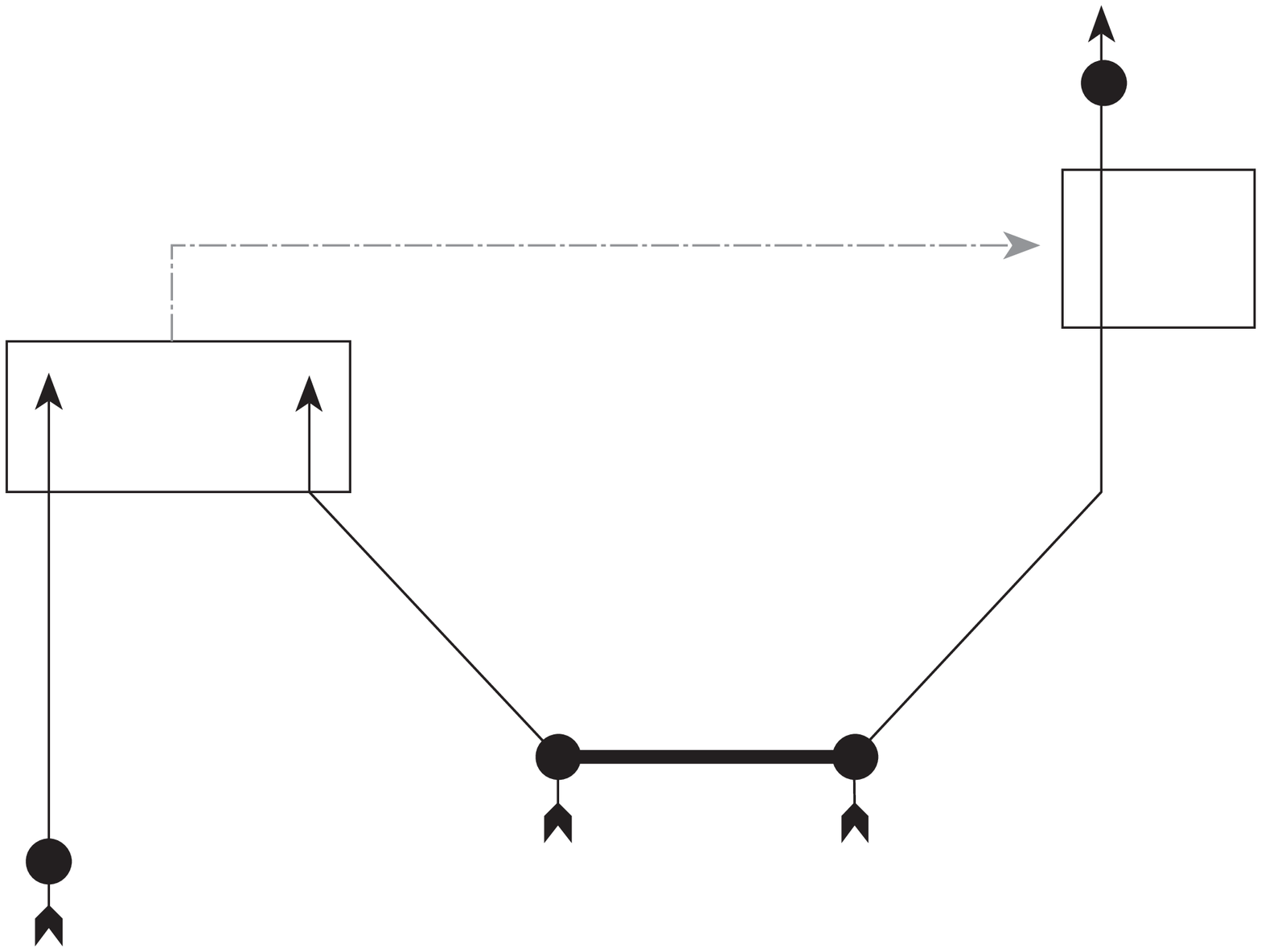,width=140pt}}       
  
\begin{picture}(140,0)   
\put(61.7,37.5){$|00\rangle\!+\!|\hspace{-0.5pt}1\hspace{-0.5pt}1\hspace{-0.5pt}\rangle$}   
\put(9,68){\small$M_{\!Bell}$}  
\put(127,86){\small$U_{\!x}$}  
\put(56.5,91){\small${x\in\mathbb{B}^2}$}  
\put(128,104.5){\small$|\phi\rangle$}  
\put(-12,17){\small$|\phi\rangle$} 
\put(165,48){\vector(0,1){40}}  
\put(167,64){${\rm time}$}  
\end{picture}  
\end{minipage}}

\vspace{-1.5mm}\noindent
The final state of $c$ proves to be $|\phi\rangle$ as well.  We will be able to
derive this fact in our abstract setting.

Since a continuous variable has been transmitted while the actual \em
classical communication \em involved only two bits, besides this \em classical
information flow \em there has to exist a \em quantum information flow\em.  The
nature of this quantum flow has been analyzed by one of the authors in
\cite{Coe1,Coe2}, building on the joint work in \cite{AC}. We recover those results
in our abstract setting (see Section 
\ref{sec:ABSETNNETW}), which also reveals additional
`fine structure'. To identify it we have to separate it from
the classical information flow. Therefore we decompose the
protocol into:
\ben
\item a {\it tree\,} with the  operations as nodes, and
  with \emph{branching} caused by the indeterminism of
measurements;
\item  a \em network \em of the operations in terms of 
the order they are applied and the subsystem
to which they apply. 
\een 

\vspace{0.8mm}\noindent{\footnotesize  
\begin{minipage}[b]{1\linewidth}   
\centering{\epsfig{figure=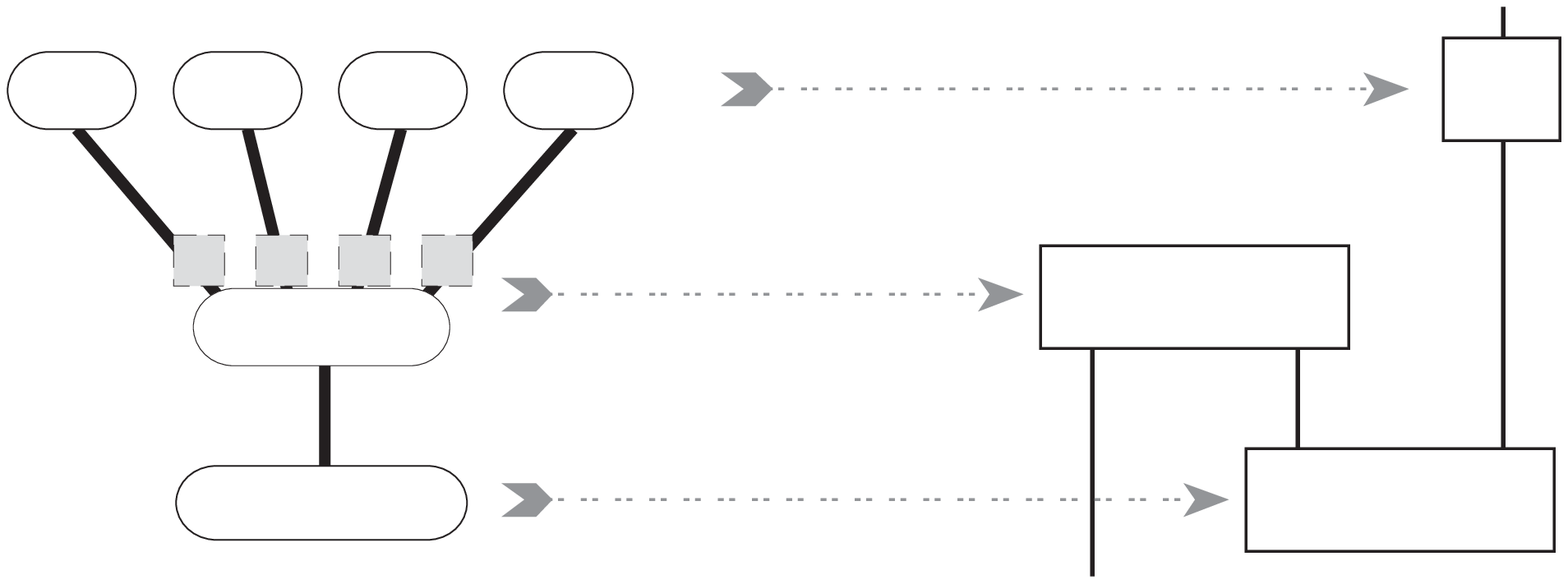,width=205pt}}       
  
\begin{picture}(205,0)   
\put(26.5,24.0){\scriptsize$|00\rangle\!+\!|\hspace{-0.5pt}1\hspace{-0.5pt}1\hspace{-0.5pt}\rangle$}     
\put(31,46.2){$M_{\!Bell}$}  
\put(3,76.9){$U_{\!00}$}   
\put(25,76.9){$U_{\!01}$}    
\put(47,76.9){$U_{\!10}$}   
\put(69,76.9){$U_{\!11}$}   
\put(22.8,56.6){\scriptsize${}_{0\hspace{-0.5pt}0}$}     
\put(33.6,56.6){\scriptsize${}_{0\hspace{-0.5pt}1}$}      
\put(44.2,56.6){\scriptsize${}_{1\hspace{-0.5pt}0}$}    
\put(55.55,56.6){\scriptsize${}_{1\hspace{-0.7pt}1}$} 
\put(194.5,75.8){$...$}   
\put(153.2,48.8){$...$}   
\put(180.7,21.5){$...$}     
\put(141,9){\normalsize$a$}  
\put(168.5,9){\normalsize$b$}  
\put(196,9){\normalsize$c$}
\end{picture}  
\end{minipage}}

\vspace{-1.8mm}\noindent
The nodes in the tree are connected to the boxes in the network by
their temporal coincidence. 
Classical communication is encoded in the tree as the dependency
of operations on the branch they are in. For each path from the
root of the tree to a leaf, by `filling in the
operations on the included nodes in the corresponding boxes of the
network', we obtain an \em entanglement network\em, that is,
a network
  
\vspace{3.4mm}\noindent{\footnotesize  
\begin{minipage}[b]{1\linewidth}  
\centering{\epsfig{figure=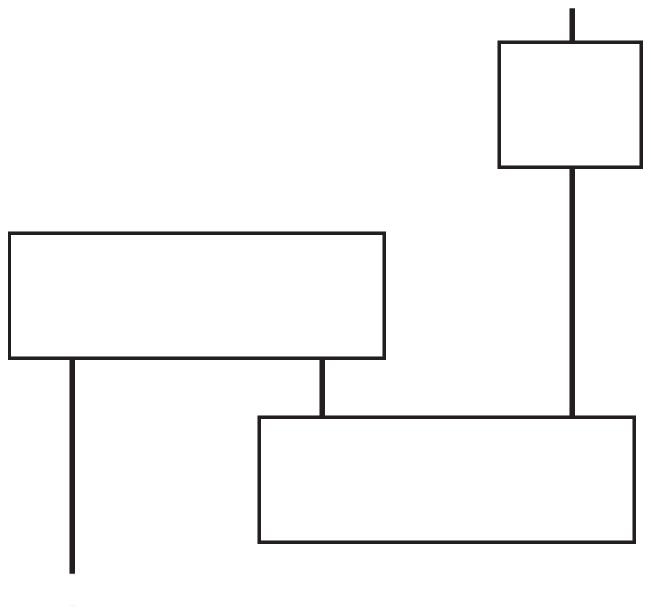,width=70pt}}      
  
\begin{picture}(70,0)   
\put(33,22.2){\scriptsize$|00\rangle\!+\!|\hspace{-0.5pt}1\hspace{-0.5pt}1
\hspace{-0.5pt}\rangle$}      
\put(16.8,42.5){${\rm P}_{\!x}$}   
\put(57.5,63.5){$U_{\!x}$}    
\put(5,8){\normalsize$a$}
\put(32.5,8){\normalsize$b$} 
\put(60,8){\normalsize$c$}
\put(95,23){\vector(0,1){40}}  
\put(97,39){${\rm time}$}  
\end{picture}  \vspace{-1mm}
\end{minipage}}
  
\vspace{0mm}\noindent
for each of the four values $x$ takes.
A component ${\rm P}_{\!x}$ of an 
observation will be referred to as an
\em observational branch\em. It will
be these networks, from which we have
removed the classical
information flow, that we will
study in Section
\ref{sec:ABSETNNETW}.  (There
is a clear analogy with the idea of
unfolding a Petri net into its set
of `processes' \cite{Petri}).
The
classical information flow will be
reintroduced in Section~9.

\subsection{Logic gate teleportation}

Logic gate teleportation \cite{Gottesman} (see also \cite{Coe1} \S 3.3) generalizes the above protocol in
that $b$ and $c$ are initially not necessarily an EPR-pair but may be in some other (not arbitrary)
entangled state $|\Psi\rangle$.  Due to this modification the final state of $c$ is not
$|\phi\rangle$ but
$|f_\Psi(\phi)\rangle$ where $f_\Psi$ is a linear map which depends on $\Psi$.  As
shown in \cite{Gottesman}, when  this construction is applied to the
situation where  $a$, $b$ and $c$ are  each a pair of qubits rather
than a single
qubit, it provides a universal quantum computational
primitive which is moreover fault-tolerant \cite{Shor} and enables
the construction of a quantum computer based on single qubit unitary operations,
Bell-base measurements and only one kind of prepared state (so-called GHZ states).
The connection between $\Psi$, $f_\Psi$ and the unitary corrections $U_{\Psi\!,x}$ will
emerge straightforwardly in our abstract setting.

\subsection{Entanglement swapping}

Entanglement swapping \cite{Swap} (see also \cite{Coe1} \S 6.2) is another
modification of the
teleportation protocol where $a$ is not in a state $|\phi\rangle$ but is a
qubit in an
EPR-pair together with an ancillary qubit $d$.  The result is that after
the protocol $c$ forms an EPR-pair
with $d$. If the measurement on $a$ and $b$ is non-destructive, we can also
perform a
unitary operation on $a$, resulting in  $a$ and $b$ also constituting
an EPR-pair.
Hence we have `swapped' entanglement:
 
\vspace{5.2mm}\noindent{\footnotesize  
\begin{minipage}[b]{1\linewidth}  
\centering{\epsfig{figure=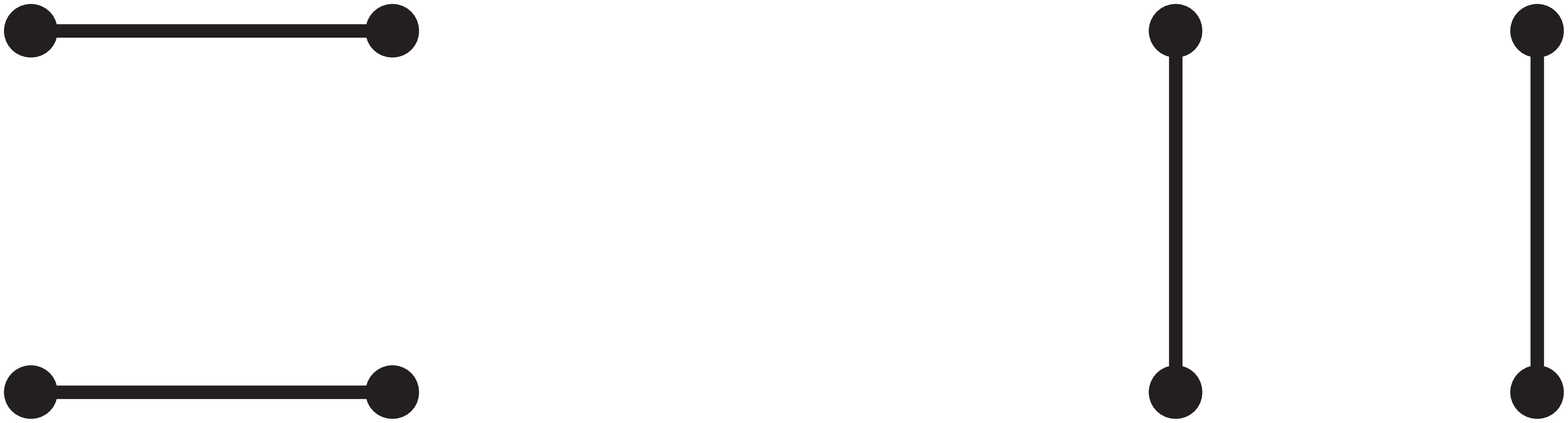,width=141.81pt}}       
   
\begin{picture}(141.81,0)   
\put(06,15.2){\tiny$|00\rangle\!\!+\!\!|\hspace{-0.5pt}1\hspace{-0.5pt}1\hspace{-0.5pt}\rangle$}      
\put(06,48.2){\tiny$|00\rangle\!\!+\!\!|\hspace{-0.5pt}1\hspace{-0.5pt}1\hspace{-0.5pt}\rangle$}      
\put(107.5,28.2){\tiny$|00\rangle\!\!+\!\!|\hspace{-0.5pt}1\hspace{-0.5pt}1\hspace{-0.5pt}\rangle$}      
\put(140.2,28.2){\tiny$|00\rangle\!\!+\!\!|\hspace{-0.5pt}1\hspace{-0.5pt}1\hspace{-0.5pt}\rangle$}      
\put(66.0,23.5){\large$\leadsto$}       
\put(-5,6){$b$}       
\put(-5,47){$a$}       
\put(38,47){$d$}       
\put(38,6){$c$}       
\put(99,6){$b$}       
\put(99,47){$a$}        
\put(142,47){$d$}       
\put(142,6){$c$}       
\end{picture}  
\end{minipage}}  
 
\vspace{-0.5mm}\noindent
In this case the entanglement networks have the shape: 

\vspace{1.8mm}\noindent{\footnotesize  
\begin{minipage}[b]{1\linewidth}  
\centering{\epsfig{figure=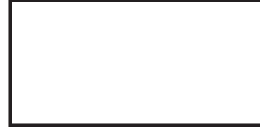,width=98pt}}       
   
\begin{picture}(98,0)   
\put(05,15.2){\scriptsize$|00\rangle\!+\!|\hspace{-0.5pt}1\hspace{-0.5pt}1\hspace{-0.5pt}\rangle$}      
\put(60.5,15.2){\scriptsize$|00\rangle\!+\!|\hspace{-0.5pt}1\hspace{-0.5pt}1\hspace{-0.5pt}\rangle$}      
\put(45.8,35.5){${\rm P}_{\!x}$}   
\put(31.5,56.5){$U_{\!x}\hspace{-1.1mm}\mbox{\rm'}$}     
\put(86.5,56.5){$U_{\!x}$}     
\put(5,2.2){\normalsize$d$}
\put(32.5,2.5){\normalsize$a$} 
\put(60,2.2){\normalsize$b$}
\put(87.5,2.5){\normalsize$c$}
\put(125,19){\vector(0,1){40}}  
\put(127,35){${\rm time}$}    
\end{picture}  
\end{minipage}}
 
\vspace{0.6mm}\noindent
Why this protocol works will again  emerge straightforwardly from our
abstract setting, as will 
generalizations of this protocol which have a much more sophisticated
compositional content (see Section
\ref{sec:ABSETNNETW}).

\Section{Compact closed categories} 

Recall that a \em symmetric monoidal category \em consists of a category {\bf C},
a bifunctorial \em tensor \em 
\[
-\otimes-:{\bf C}\times{\bf C}\to{\bf C}\,,
\]
a \em unit \em object
${\rm I}$ and natural isomorphisms 
\[
\lambda_A : A \simeq {\rm I}\otimes A\quad\quad\quad\quad\quad\ \ \rho_A: A \simeq
A\otimes{\rm I}
\]
\[
\alpha_{A,B,C}:A\otimes(B\otimes C)\simeq (A\otimes B)\otimes C\vspace{0.5mm} 
\] 
\[
\sigma_{A,B}:A\otimes B\simeq B\otimes A
\]
which satisfy certain coherence conditions \cite{MacLane}.
A category {\bf C} is \em $*$-autonomous \em \cite{Barr} if
it is symmetric monoidal, and comes equipped with a full and
faithful functor 
\[
(\ )^*:{\bf C}^{op}\to{\bf C}
\]
such that a bijection
\[ 
{\bf C}(A\otimes B,C^*)\simeq {\bf C}(A,(B\otimes C)^*)
\]
exists which is natural in all variables. 
Hence a $*$-autonomous category is closed, with
\[
{A\multimap B:=(A\otimes B^*)^*}\,.
\]
These $*$-autonomous categories
provide a categorical semantics for the multiplicative
fragment of linear logic \cite{Seely}.

A \em compact closed category \em \cite{Kelly} is a
$*$-autonomous category with a self-dual tensor,
i.e.~with natural isomorphisms
\[
u_{A,B}:(A\otimes B)^*\simeq A^*\otimes B^* \qquad u_{\II} : \II^* \simeq
\II\,.
\] 
It follows that 
\[
A\multimap B\simeq A^*\otimes B\,.
\] 

\subsection{Definitions and examples}

A very different definition arises when one considers a symmetric monoidal
category as a one-object bicategory. In this context,
compact closure simply means that every object $A$, qua 1-cell of the
bicategory, has an adjoint \cite{KL}. 
\begin{definition}[Kelly-Laplaza]\label{def:compclos}\em 
A \em compact closed category \em is a symmetric monoidal
category in which to each object $A$ a \em dual object \em  
$A^*$, a \em unit \em 
\[
\eta_A:{\rm I}\to A^*\otimes A
\]
 and a \em  
counit \em 
\[
\epsilon_A:A\otimes A^*\to {\rm I}
\]
are assigned in
such a way that the diagram
\begin{diagram}  
A&\rTo^{\rho_A}&A\otimes{\rm
I}&\rTo^{1_A\otimes\eta_A}&A\otimes(A^*\otimes
A)\\ 
\dTo^{1_A}&&&&\dTo^{\alpha_{A,A^*\!\!,A}}\\ 
A&\lTo_{\lambda_A^{-1}}&{\rm I}\otimes A&\lTo_{\epsilon_A\otimes
1_A}&(A\otimes A^*)\otimes A
\end{diagram}
and the dual one for $A^*$ both commute.  
\end{definition} 
 
The monoidal categories $({\bf Rel},\times)$ of sets, relations
and cartesian product and $({\bf FdVec}_\mathbb{K},\otimes)$ of finite-dimensional vector spaces over a field
$\mathbb{K}$, linear maps and tensor product
are both compact closed.  In $({\bf Rel},\times)$, taking a one-point set $\{ * \}$ as the unit for $\times$, and writing $R^{\cup}$ for the converse of a relation $R$:
\[
\eta_X=\epsilon_X^{\cup}=\{(*,(x,x))\mid x\in X\}\,.
\] 
The unit and counit in $({\bf FdVec}_\mathbb{K},\otimes)$ are\vspace{-0.5mm}
\beqa 
&&\hspace{-6mm}\eta_V:\mathbb{K}\to V^*\otimes V::1\mapsto\sum_{i=1}^{i=n}\bar{e}_i\otimes
e_i\\ &&\hspace{-6mm}\epsilon_V:V\otimes
V^*\to\mathbb{K}::e_i\otimes\bar{e}_j\mapsto \bar{e}_{j}( e_{i})
\eeqa 
where $n$ is the dimension of $V$, $\{e_i\}_{i=1}^{i=n}$ is a basis of
$V$ and $\bar{e}_i$ is the linear functional in $V^*$ determined by
$\bar{e}_{j}( e_{i}) = \delta_{ij}$.

\begin{definition}\label{def:name}\em
The \em name \em $\uu f\uuu$ and the \em coname \em $\dd f\ddd$ of a morphism $f:A\to B$ in a compact
closed category are
\begin{diagram} 
A^*\!\!\otimes\! A&\rTo^{1_{A^*}\!\!\otimes\! f}&A^*\!\otimes\! B&&&{\rm I}\\
\uTo^{\eta_A}&\ruTo_{\uu f\uuu}&&&\ruTo^{\dd f\ddd}&\uTo_{\epsilon_B} \\     
{\rm I}&&&A\!\otimes\! B^*&\rTo_{f\!\otimes\! 1_{B^*}}&B\!\otimes\! B^*&&
\end{diagram}
\end{definition}

For $R\in{\bf Rel}(X,Y)$ we have
\beqa
&&\hspace{-6mm}\uu R\uuu=\{(*,(x,y))\mid xRy,x\in X, y\in Y\}\\
&&\hspace{-6mm}\dd R\ddd=\{((x, y),*)\mid xRy,x\in X, y\in Y\}
\eeqa

\noindent 
and for $f\in {\bf FdVec}_\mathbb{K}(V,W)$ with $(m_{ij})$ the matrix of
$f$ in bases
$\{e_i^V\}_{i=1}^{i=n}$ and $\{e_j^W\}_{j=1}^{j=m}$ of $V$ and $W$
respectively:
\beqa
&&\hspace{-6mm}\uu f\uuu:\mathbb{K}\to V^*\otimes
W::1\mapsto\sum_{i,j=1}^{\!i,j=n,m\!}m_{ij}\cdot
\bar{e}_i^V\otimes e_j^W\\ &&\hspace{-6mm}\dd f\ddd:V\otimes
W^*\to\mathbb{K}::e_i^V\otimes\bar{e}_j^W\mapsto m_{ij} .
\eeqa

\subsection{Some constructions}

Given $f:A\to B$ in any compact closed category ${\bf C}$  we can define $f^*:B^*\to A^*$ as follows
\cite{KL}:
\begin{diagram}  
B^*&\rTo^{\lambda_{B^*}}&{\rm I}\otimes B^*&\rTo^{\eta_A\otimes 1_{B^*}}&A^*\otimes A\otimes
B^*\\ 
\dTo^{f^*}&&&&\dTo^{1_{A^*}\!\otimes f\otimes 1_{B^*}}\\   
A^*&\lTo_{\rho_{A^*}^{-1}}&A^*\otimes {\rm I}&\lTo_{1_{A^*}\otimes \epsilon_B}&A^*\otimes
B\otimes B^*
\end{diagram}
This operation $(\ )^*$ is functorial and makes Definition \ref{def:compclos} coincide
with the one given at the beginning of this section. It then follows by 
\[
{\bf C}(A\otimes B^*,{\rm I})\simeq{\bf C}(A,B)\simeq{\bf C}(I,A^*\otimes B)
\]
that every morphism
of type $\II\!\to\!   A^*\!\otimes B$ is the name of some morphism of type ${A\to B}$ and every
morphism of type ${A\otimes B^*\!\to{\rm I}}$ is the coname of some morphism of
type ${A\to B}$.  In the case of the unit and the counit we have
\[
\eta_A={\uu 1_A\uuu}\quad\quad{\rm and}\quad\quad\epsilon_A={\dd 1_A\ddd}\,.  
\]
For $R\in{\bf Rel}(X,Y)$ the dual is the converse,
$R^*=R^{\cup}\in{\bf Rel}(Y,X)$, and for $f\in{\bf FdVec}_\mathbb{K}(V,W)$, the dual is
\[
f^*:W^*\to V^*::\,\phi \mapsto\,\phi\circ f\,.
\]

In any compact closed category, there is a natural isomorphism
$d_A:A^{**}\simeq A$.

The following holds by general properties of adjoints and the fact that the tensor is symmetric \cite{KL}.

\begin{proposition}\label{prop:triqcc}
In a compact closed category ${\bf C}$ we have
\begin{diagram}
\II&\rTo^{\eta_{A^*}}&A^{**}\otimes A^*&&A^*\otimes
A&\rTo^{\sigma_{A^*\!\!,A}}&A\otimes A^*\\
\dTo^{\eta_A}&&\dTo~{d_A\otimes 1_{A^*}\!\!\!}&&\dTo~{1_{A^*}\otimes d^{-1}_A}&&\dTo_{\epsilon_A}\\
A^*\otimes A&\rTo_{\sigma_{A^*\!\!,A}}&A\otimes A^*&&A^*\otimes
A^{**}&\rTo_{\epsilon_{A^*}}&\II
\end{diagram}
for all objects $A$ of ${\bf C}$.
\end{proposition}

\subsection{Key lemmas}

The following Lemmas constitute the core of our
interpretation of entanglement in compact closed categories.
 
\begin{lemma}[absorption]\label{lm:precompos} For
\begin{diagram}
A&\rTo^{f}&B&\rTo^{g}&C
\end{diagram}
we have that
\[
(1_{A^*}\!\!\otimes g)\circ \uu f\uuu=\uu g\circ f\uuu.
\]
\end{lemma}
\bpf
Straightforward by Definition \ref{def:name}.
\hfill\endproof

\begin{lemma}[compositionality]\label{lm:compos} For
\begin{diagram}
A&\rTo^{f}&B&\rTo^{g}&C
\end{diagram}
we have that
\[
\lambda^{-1}_C\circ (\dd f\ddd\otimes 1_C)\circ(1_A\otimes\uu
g\uuu)\circ\rho_A=g\circ f\,.
\]
\end{lemma}
\bpf
See the diagram in the appendix to this paper which uses bifunctoriality and naturality of
$\rho$ and
$\lambda$.
\hfill\endproof

\begin{lemma}[compositional CUT]\label{lm:CUT}   
For
\begin{diagram}
A&\rTo^{f}&B&\rTo^{g}&C&\rTo^{h}&D
\end{diagram}
we have that
\[ 
\hspace{-1.5mm}(\rho^{-1}_A\!\otimes 1_{D^*}\!)\circ(1_{A^*}\!\otimes\dd g\ddd\otimes\!
1_D)\circ({\uu f\uuu}\otimes{\uu  h\uuu})\circ\rho_I =\uu h\circ g\circ f\uuu.
\hspace{-1.5mm}
\]
\end{lemma}
\bpf
See the diagram in the appendix to this paper which uses Lemma \ref{lm:compos} and naturality of
$\rho$ and
$\lambda$.
\hfill\endproof\newline

On the right hand side of Lemma \ref{lm:compos} we have $g\circ f$, that is, we
first apply $f$ and then $g$, while on the left hand side we first apply  the coname of $g$, and then
the coname of $f$.  In Lemma~\ref{lm:CUT} there is a similar,
seemingly `acausal' inversion of the order of application, as $g$ gets
inserted between $h$ and $f$.

For completeness we  add the following `backward' absorption lemma, which again involves a
reversal of the composition order.

\begin{lemma}[backward absorption]\label{lm:precompos2} For 
\begin{diagram}
C&\rTo^{g}&A&\rTo^{f}&B
\end{diagram}
we have that
\[
(g^*\otimes 1_{A^*}\!)\circ \uu f\uuu=\uu f\circ g\uuu. 
\]
\end{lemma}
\bpf
This follows by unfolding the definition of $g^*$, then using naturality of $\lambda_{A^*}$, $\lambda_\II=\rho_\II$, and finally Lemma \ref{lm:CUT}.
\hfill\endproof\newline

\noindent The obvious analogues of Lemma \ref{lm:precompos} and  \ref{lm:precompos2} for conames also hold.

\Section{Abstract entanglement networks}\label{sec:ABSETNNETW}   

We claim that Lemmas \ref{lm:precompos}, \ref{lm:compos} and \ref{lm:CUT} capture the quantum
information flow in the (logic-gate) teleportation and entanglement swapping protocols.  
We shall provide a full interpretation of finitary quantum
mechanics in Section \ref{sec:absquantprot} but for now the following rule
suffices:
\bit
\item We interpret \emph{preparation} of an entangled state as a \emph{name} and an \emph{observational branch} as a \emph{coname}. 
\eit
For an entanglement network of teleportation-type shape (see the
picture below) applying Lemma \ref{lm:compos} yields
\[
U\circ\left(\lambda^{-1}_C\circ (\dd f\ddd\otimes
1)\right)\circ\left((1\otimes\uu g\uuu)\circ\rho_A\right)
=U\circ g\circ f\,.
\]

\noindent
Note that the quantum information seems to flow `following the line' while being
acted on by the functions whose name or coname labels the boxes
(and this fact remains valid for much more complex networks
\cite{Coe1}).

\vspace{1.3mm}\noindent{   
\begin{minipage}[b]{1\linewidth}  
\centering{\epsfig{figure=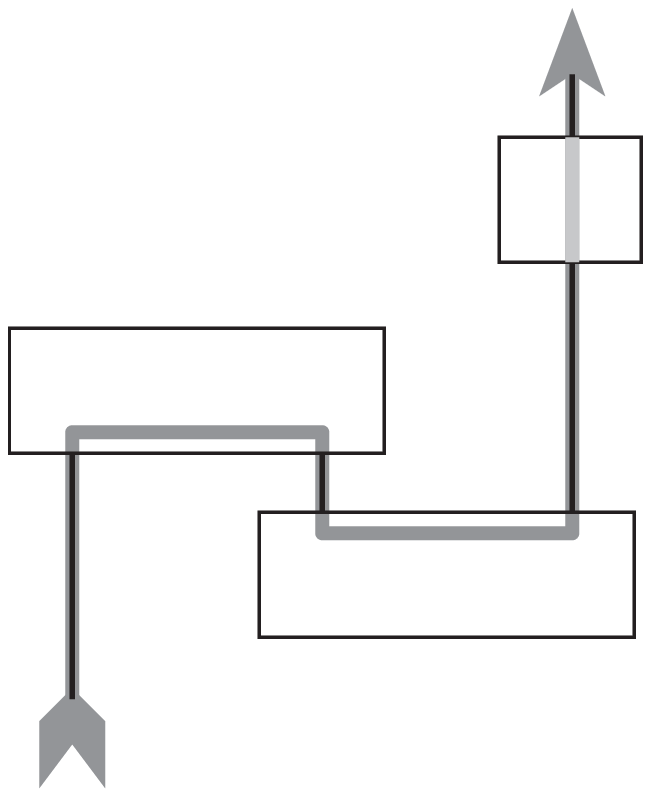,width=70pt}}        
  
\begin{picture}(70,0)   
\put(41.2,31.4){$\uu g\uuu$}       
\put(13.5,54.6){$\dd f\ddd$}   
\put(59.0,73.5){$U$}    
\put(95,38){\vector(0,1){40}}  
\put(97,54){\footnotesize ${\rm time}$}  
\end{picture}  
\end{minipage}} 
 
\vspace{-4.6mm}\noindent
Teleporting the input requires
$U\circ g\circ f=1_A$ --- we assume all functions have type $A\to A$.
Logic-gate teleportation of $h:A\to B$ requires $U\circ g\circ f=h$.

We 
calculate this explicitly in {\bf Rel}. For initial state
$x\in X$ after preparing 
\[
{\uu S\uuu\subseteq\{*\}\times(Y\times Z)}
\]
we obtain 
\[
\{x\}\times \{(y,z)\mid *\,\uu S\uuu(y,z)\}
\]
as the state
of the system. For observational branch
\[
\dd R\ddd\subseteq (X\times Y)\times \{*\}
\] 
we have that $z\in Z$ is the output iff $\dd R\ddd\times 1_Z$
receives  an input $(x,y,z)\in X\times Y\times Z$
such that $(x,y)\dd R\ddd\,*\,$.
Since 
\[
*\,\uu S\uuu(y,z)\Leftrightarrow ySz\quad {\rm and}\quad (x,y)\dd
R\ddd\,*\Leftrightarrow xRy
\]
we indeed obtain $x(R;S)z$.
This illustrates that the compositionality
is due to a mechanism of imposing constraints between the components
of the tuples.

In ${\bf FdVec}_\mathbb{C}$   
the vector space of all linear maps of type $V\to W$ is $V\multimap W$ and
hence by 
\[
{V^*\otimes W\simeq V\multimap W}
\]
we have a bijective
correspondence between linear maps
${f:V\to W}$ and vectors $\Psi\in V^*\otimes W$ (see also
\cite{Coe1,Coe2}):
\[
\sqrt{2}\cdot \Psi_f=\uu f\uuu(1)\quad\quad\quad\quad\quad\dd
f\ddd=\langle\sqrt{2}\cdot\Psi_f|-\rangle\,.
\]
In particular we have for the Bell base:
\[
\sqrt{2}\cdot b_i=\uu \beta_i\uuu(1)\quad\quad\quad\quad\quad\dd
\beta_i\ddd=\langle \sqrt{2}\cdot b_i|-\rangle\,.
\]  
Setting 
\[
g:=\beta_1=1_V\,,\quad f:=\beta_i\,,\quad U:=\beta_i^{-1}
\]
indeed yields 
\[
\beta_i^{-1}\circ 1_A\circ\beta_i=1_A\,,
\] 
which expresses the correctness
of the teleportation protocol along each branch.

Setting $g:=h$ and $f:=\beta_i$ for logic-gate teleportation requires
$U_i$ to satisfy
$U_i\circ h\circ \beta_i=h$ that is 
\[
{h\circ
\beta_i=U^\dagger\circ h}
\]
(since $U$ has to be unitary).
Hence we have derived the laws of logic-gate teleportation --- one should
compare this calculation to the size of the calculation in
Hilbert space.

Deriving the swapping protocol using
Lemma \ref{lm:precompos} and Lemma \ref{lm:CUT} proceeds analogously
to the derivation of the teleportation protocol. We obtain
two distinct flows due to the fact that a non-destructive
measurement is involved.

\vspace{1.8mm}\noindent{\footnotesize  
\begin{minipage}[b]{1\linewidth}  
\centering{\epsfig{figure=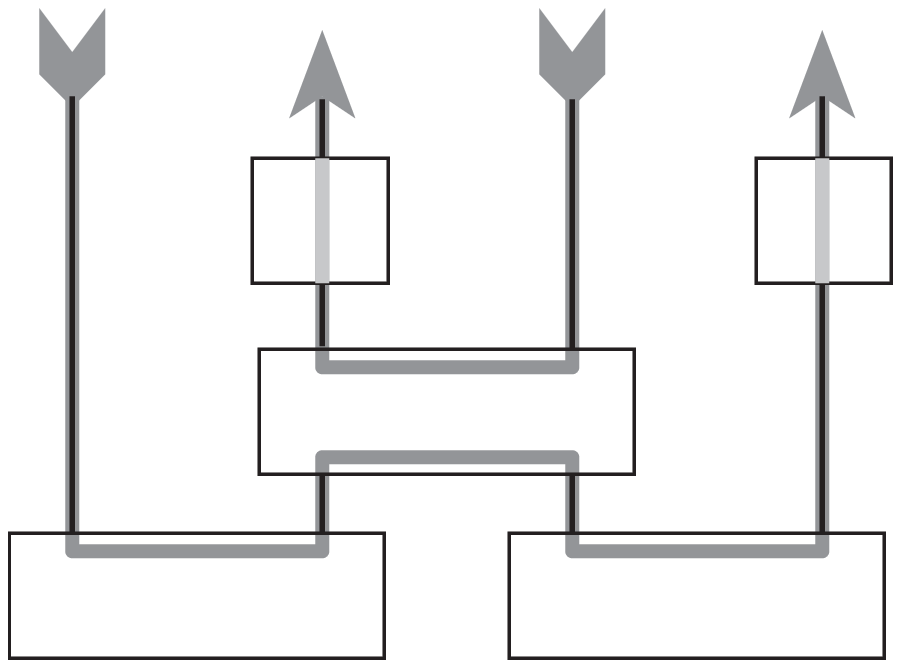,width=98pt}}       
   
\begin{picture}(98,0)   
\put(15,13.2){$\uu 1\uuu$}      
\put(70.5,13.2){$\uu 1\uuu$}      
\put(30.3,35.5){$\uu\gamma_i\!\uuu\!\circ\!\dd\beta_i\ddd$}   
\put(29.1,56.5){$\gamma_i^{-\!1}$}    
\put(84.5,56.5){$\beta_i^{-\!1}$}      
\put(125,24){\vector(0,1){40}}  
\put(127,40){${\rm time}$}   
\end{picture}  
\end{minipage}}
 
\vspace{-3mm}
How $\gamma_i$ has to relate to $\beta_i$ such that they make up a true projector
will be discussed in Section \ref{sec:absquantprot}.

For a general entanglement
network of the swapping-type (without unitary correction and
observational branching) by Lemma
\ref{lm:CUT} we obtain the following `reduction':\vspace{-0mm}

\vspace{1.8mm}\noindent{  
\begin{minipage}[b]{1\linewidth}  
\centering{\epsfig{figure=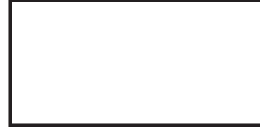,width=196pt}}       
   
\begin{picture}(196,0)   
\put(120.0,23.5){\large$\leadsto$}      
\put(13.5,16.2){$\uu f\uuu$}      
\put(69.5,16.2){$\uu h\uuu$}      
\put(41.8,37.5){$\dd g\ddd$}   
\put(155.35,23.5){$\uu h\!\circ\! g\!\circ\! f\uuu$}    
\end{picture}  
\end{minipage}}
 
\vspace{-3mm}\noindent
This picture, and the underlying algebraic property expressed by Lemma 
3.5, is in fact directly related to \emph{Cut-Elimination} in the
logic corresponding to compact-closed categories. If one turns the
above picture upside-down, and interprets names as Axiom-links and
conames as Cut-links, then one has a normalization rule for
proof-nets. This  perspective is 
developed in \cite{Ross}.

\Section{Biproducts}\label{sec:biprods} 

Biproducts have been studied as part of the structure of Abelian
categories. For further details, and proofs of the general results
we shall cite in this Section, see e.g. \cite{Mitchell}.

A \emph{zero object}  in a category is one which is both
initial and terminal. If $\Zero$ is a zero object, there is an 
arrow
\begin{diagram}
\hspace{-8mm}0_{A,B}:A&\rTo&\Zero &\rTo &B
\end{diagram}
between any pair of objects $A$ and $B$. Let $\CC$ be a category with
a zero object and binary products and coproducts.
Any arrow 
\[
A_1 \coprod A_2 \rightarrow A_1 \prod A_{2}
\]
can be written uniquely
as a matrix $(f_{ij})$, where $f_{ij} : A_{i} \rightarrow A_j$.
If the arrow 
\[ \left( \begin{array}{cc}
1 & 0 \\
0 & 1
\end{array} \right) \]
is an isomorphism for all $A_1$, $A_2$, then we say that $\CC$ has
\emph{biproducts}, and write $A \oplus B$ for the biproduct of $A$ and 
$B$. 
\begin{proposition}[Semi-additivity]
If $\CC$ has biproducts, then we can define an operation of addition
on each hom-set $\CC (A, B)$ by
\begin{diagram} 
A&\rTo^{f+g}&B\\
\dTo^\Delta&&\uTo_\nabla\\
A\oplus A&\rTo_{f\oplus g}&B\oplus B
\end{diagram}
for  $f,g:A\to B$, where 
\[
\Delta=\langle 1_A,1_A\rangle\qquad {\rm and}\qquad
\nabla=[1_B,1_B]
\]
are respectively the diagonal and 
codiagonal.
This operation is associative and commutative, with $0_{AB}$ as an
identity. Moreover, composition is bilinear with respect to this
additive structure. Thus $\CC$ is enriched over abelian monoids.
\end{proposition}
 
\begin{proposition}
If $\CC$ has biproducts, we can choose projections $p_1$, $\ldots$, $p_n$ and
injections $q_1$, $\ldots$, $q_n$ for each $\bigoplus_{k=1}^{k=n}A_k$ satisfying\vspace{-3mm}
\[ p_j \circ q_i = \delta_{ij} \qquad{\rm and}\qquad \sum_{k=1}^{k=n}q_k \circ p_k=
1_{\bigoplus_{k}\!A_k} \] 
where $\delta_{ii} = 1_{A_i}$, and $\delta_{ij} = 0_{A_{i}, A_{j}}$, $i \neq j$.
\end{proposition}

%

\begin{proposition}[Distributivity of $\otimes$ over $\oplus$]\label{distributivity}
In monoidal closed categories there are natural isomorphisms 
\[
\tau_{A,B,C}:A\otimes(B\oplus C)\simeq (A\otimes B)\oplus(A\otimes C)\,,
\]
explicitly,
\[
\tau_{A,\cdot,\cdot}\!\!=\langle 1_A\otimes p_1,1_A\otimes p_2\rangle\quad\ \ 
\tau_{A,\cdot,\cdot}^{-1}\!\!=[ 1_A\otimes q_1,1_A\otimes q_2]\,.
\]
A left distributivity isomorphism 
\[ \upsilon_{A, B, C} : (A \oplus B) \otimes C
\simeq (A \otimes C) \oplus (A \otimes C) \]
can be defined similarly.
\end{proposition}

\begin{proposition}[Self-duality of $\oplus$ for $(\ )^*$]
In any compact closed category there are natural isomorphisms
\[
\nu_{A,B}:(A\oplus B)^*\simeq A^*\oplus B^*\qquad
\nu_\II:\Zero^*\simeq\Zero\,.
\]
\end{proposition}
Writing $n\cdot X$ for $\bigoplus_{i=1}^{i=n}X$ it follows by
self-duality of the tensor unit
$\II$ that
\[ 
\nu^{-1}_{{\rm I},\ldots,{\rm
I}}\circ\left(n\cdot{u}_{\rm
I}\right)\ :\,n\cdot{\rm I}\simeq\left(n\cdot{\rm
I}\right)^{*}\,.
\]

\paragraph{Matrix representation.} 
We can write any arrow of the form
$f:A\oplus B\to C\oplus D$ as a matrix
\[
M_f:=
\left(
\begin{array}{cc}
p_1^{C,D}\!\!\circ f\circ q_1^{A,B} & p_1^{C,D}\circ f\circ q_2^{A,B}\\
p_2^{C,D}\circ f\circ q_1^{A,B} & p_2^{C,D}\circ f\circ q_2^{A,B}\\
\end{array}
\right).
\]
The sum $f+g$ of such morphisms corresponds to the matrix sum $M_f+M_g$ and composition $g\circ f$ corresponds to matrix
multiplication $M_g\cdot M_f$. Hence  categories with biproducts admit a matrix
calculus.
 
\paragraph{Examples.} The categories $({\bf
Rel},\times,+)$ where the biproduct is the disjoint union and
$({\bf FdVec}_\mathbb{K},\otimes,\oplus)$ where the biproduct is the
direct sum are examples of compact closed categories with biproducts.  
More generally, the category of relations for a regular category with
stable disjoint coproducts; the category of finitely generated projective modules
over a commutative ring; the category of finitely generated free semimodules over a
commutative semiring; and the category of free semimodules over a
complete commutative semiring are all compact closed with biproducts. 
Compact closed categories with biproducts, with additional assumptions
(e.g. that the category is abelian) have been studied in the
mathematical literature on \emph{Tannakian categories} \cite{Deligne}.
They have also arisen in a Computer Science context in the first
author's work on Interaction Categories \cite{Abramsky}.

\Section{Scalars} 
In any compact closed category we shall call endomorphisms $s : \II \rightarrow
  \II$ \emph{scalars}. As observed in \cite{KL}, in any monoidal
  category $\CC$, the endomorphism
  monoid $\CC (\II , \II )$ is commutative. Any scalar $s$ induces a natural transformation $s_A 
  : A \rightarrow A$ by
\begin{diagram}
A & \rTo^{\!\!\!\!\lambda} & \II \otimes A & \rTo^{s \otimes 1_A} & \II
\otimes A & \rTo^{\lambda^{-1}\!\!\!\!} & A\,.
\end{diagram} 
Here naturality means that all morphisms `preserve scalar
multiplication'. We write $s \sdot f$ for $f \circ s_A$, where $s$ is a scalar and $f : 
A \rightarrow B$. If $\CC$ moreover has biproducts, 
the scalars $\CC (\II , \II )$ form a commutative semiring. 

\paragraph{Examples.} In $\mathbf{FdVec}_{\mathbb{K}}$, linear maps $s:\mathbb{K} \to \mathbb{K}$ are uniquely
determined by the image of $1$, and hence correspond biuniquely to
elements of $\mathbb{K}\,$; composition and addition of these maps
corresponds to multiplication and addition of 
scalars. Hence in  $\mathbf{FdVec}_{\mathbb{K}}$ the commutative semiring of scalars is the field $\mathbb{K}$.
In $\mathbf{Rel}$, there are just two scalars, corresponding to the classical truth values.
Hence in $\mathbf{Rel}$ the commutative semiring of scalars is the Boolean semiring $\{ 0, 1 \}$.

\Section{Strong compact closure}
In any compact closed category
$\CC$, there is a natural isomorphism $A \simeq A^{**}$. It will be
notationally convenient to assume that $(\ )^*$ is strictly
involutive, so that this natural isomorphism is the identity. The
following definition allows the key example of (\em complex\em)
Hilbert spaces to be  accomodated in our
setting.

\begin{definition}\em
A compact closed category $\CC$ is \emph{strongly compact closed}
if the assignment on objects ${A \mapsto A^*}$ extends to a 
\emph{covariant} functor, with action on morphisms $f_* : A^* \to
B^*$ for $f : A \to B$, such that
\[ f_{**} = f \qquad \quad
(f_* )^* = (f^* )_* : B \to A\,. 
\]
\end{definition}

\paragraph{Examples.} Any compact closed category such as
$\mathbf{Rel}$, in which $(\ )^*$ is
the identity on objects, is trivially strongly compact closed (we just 
take $f_* \!:= f$). The
category of finite-dimensional  real inner product spaces and linear
maps offers another example of this situation, where we take $A =
A^*$, and define
\[ 
\epsilon_A : \phi \otimes \psi \mapsto \langle \phi \mid \psi
\rangle\, . 
\]

Our main intended example, $\FdHilb$, the category of
finite-dimensional Hilbert spaces and linear maps, exhibits this
structure less trivially, since the conjugate-linearity in the first
argument of the inner product prevents us from proceeding as for real 
spaces. Instead, we define $\HH^*$ as follows.
The additive abelian group of
vectors   in $\HH^*$ is the same as in $\HH$. Scalar
multiplication and the inner product are 
\[ \alpha \sdot_{\HH^*} \phi := \bar{\alpha} \sdot_{\HH} \phi \qquad 
\langle \phi \mid \psi \rangle_{\HH^*} := \langle \psi \mid \phi
\rangle_{\HH} \]
where $\bar{\alpha}$ is the complex conjugate of $\alpha$.
The  covariant action is then just $f_* = f$. 
Note that the identity map from $\HH$ to $\HH^*$ is a conjugate-linear 
isomorphism, but \emph{not} linear --- and hence does not live in the
category $\FdHilb$! Importantly, however, $\HH$ and $\HH^*$ have the
same orthonormal bases. Hence we can define
\[ 
\eta_{\HH} : 1 \mapsto \sum_{i=1}^{i=n} e_i \otimes e_i 
\qquad  \ \
\epsilon_{\HH} : \phi \otimes \psi\mapsto \langle \psi \mid \phi\rangle_{\HH}
\]
where $\{ e_i \}_{i=1}^{i=n}$ is an orthonormal basis of $\HH$.

\subsection{Adjoints, unitarity and inner products}

Each morphism in a strongly compact closed category admits an adjoint in the following sense. 

\begin{definition}\em
We set 
\[
f^{\dagger} := (f_* )^* = (f^* )_*\,,
\]
and call this the
\emph{adjoint} of $f$. 
\end{definition} 

\begin{proposition}
The assignments $A \mapsto A$ on objects, and $f \mapsto f^{\dagger}$
on morphisms, define a
contravariant involutive functor: 
\[ (f\circ g)^\dagger=g^\dagger\circ f^\dagger \qquad 1^{\dagger} = 1 \qquad 
f^{\dagger
\dagger}\! = f . \]
\end{proposition}
In $\FdHilb$ and real inner product spaces, $f^\dagger$ is the usual adjoint of a 
linear map. In \textbf{Rel}, it is relational converse. 

\begin{definition}\em
An isomorphism $U$ is called \em unitary \em if its adjoint is its 
inverse ($U^\dagger=U^{-1}$).
\end{definition}

\begin{definition}\em
Given $\psi,\phi : \II \to A$ we define  their \em
abstract inner product \em $\langle\psi\mid\phi\rangle$  as 
\begin{diagram}
\ \II \!& \rTo^{\rho_{\II}} & \!\II\! \otimes\! \II\!& \rTo^{1_\II \!\otimes\! u_{\II}} & \II\!
\otimes\!
\II^*\!\! & \rTo^{\phi \!\otimes\! \psi_*} & \!A \!\otimes\! A^*\!\! & \rTo^{\epsilon_A} & \!\II. 
\end{diagram}
\end{definition}
In $\FdHilb$, this definition coincides with the usual inner product.
In {\bf Rel} we have for $x,y\subseteq\{*\}\times X$:
\[
\langle x\mid y\rangle= 1_{\II}, \quad x \cap y \neq \varnothing \qquad
\langle x\mid y\rangle= 0_{\II}, \quad x \cap y = \varnothing .
\]

\begin{lemma}\label{lm:expldaggaer}
\label{adjlemma}
If $\psi : \II \to A$, then $\psi^{\dagger}$ is given by
\begin{diagram}
\ A \!& \rTo^{\rho_A} & \!A\! \otimes\! \II\!& \rTo^{1_A \!\otimes\! u_{\II}}
& A\!
\otimes\!
\II^*\!\! & \rTo^{1_A\! \otimes\! \psi_*} & \!A \!\otimes\! A^*\!\! & \rTo^{\epsilon_A} & \!\II. 
\end{diagram}
\end{lemma}
\bpf
See the diagram in the appendix to this paper, which  uses Proposition \ref{prop:triqcc} twice (with $d_A= 1_A$).
\hfill\endproof

\begin{proposition}\label{pro:inprod}
For $\langle \psi \mid \phi \rangle$ as defined above we have 
\[ 
\langle \psi \mid \phi \rangle = \psi^{\dagger}\circ \phi\,. 
\]
\end{proposition}
\bpf
Using bifunctoriality of tensor and naturality of $\rho$, it is easy to see that $\langle \psi  \mid \phi \rangle$ can be
written as
\[
\II  \rTo^{\phi}  A  \rTo^{\rho_A}  A\! \otimes\! \II  \rTo^{1_A \otimes u_{\II}}  A\! \otimes\!
\II^*\!  \rTo^{1_A \otimes \psi_*}  A\! \otimes\! A^*\!  \rTo^{\epsilon_A}\!  \II\,.
\]
We now apply Lemma~\ref{adjlemma} to conclude. 
\hfill\endproof\newline

\begin{proposition}\label{pr:adjIn}
For
\[
\psi:\II\to A\qquad \phi:\II\to B\qquad f:B\to A 
\]
we have 
\[
\langle f^\dagger\circ\psi\mid\phi\rangle_B=\langle \psi\mid f\circ\phi\rangle_A\,.
\]
\end{proposition}
\bpf
By Proposition \ref{pro:inprod},
\[
\langle
f^\dagger\!\circ\psi\mid\phi\rangle=(f^\dagger\!\circ\psi)^\dagger\!\circ\phi=
\psi^\dagger\!\circ f\circ\phi=\langle \psi\mid 
f\circ\phi\rangle. \ \ 
\]
\vspace{-11.0mm} 

\hfill\endproof

\begin{proposition}\label{pr:UnIn}
Unitary morphisms $U:A\to B$ preserve the inner product, that is  
for all $\psi,\phi:\II\to A$ we have 
\[
\langle U\circ\psi\mid U\circ\phi\rangle_B=
\langle \psi\mid \phi\rangle_A\,.
\]
\end{proposition}
\bpf
By Proposition \ref{pr:adjIn},
\beqa
\langle U\circ\psi\mid U\circ\phi\rangle_B=
\langle U^\dagger\!\circ U\circ\psi\mid \phi\rangle_A=
\langle \psi\mid \phi\rangle_A.
\eeqa
\vspace{-11.0mm} 

\hfill\endproof

\subsection{Bras and kets}

By Proposition \ref{pr:adjIn} we can interpret the Dirac notation
(e.g.~\cite{Nielsen}) in our setting. For morphisms
\[
\psi:\II\to A\qquad \phi:\II\to B\qquad f:B\to A 
\]
define
\[
\langle \psi \mid f \mid \phi \rangle:=\langle
f^\dagger\circ\psi\mid\phi\rangle_B=\langle \psi\mid f\circ\phi\rangle_A .
\]
By Proposition \ref{pro:inprod},
\[ 
\langle \psi \mid f \mid \phi \rangle = \psi^{\dagger} \circ f \circ \phi\,. 
\]

\subsection{Strong compact closure and biproducts}

\begin{proposition}
If $\CC$ has biproducts, $(\
)^{\dagger}$ preserves them and hence is additive:
\[ 
(f + g)^{\dagger} = f^{\dagger} + g^{\dagger} \qquad\qquad 0^{\dagger}_{A,B} = 0_{B,A}\,. 
\]
\end{proposition}

If a category is both strongly compact
closed and has biproducts, the adjoint acts as an  involutive automorphism  on the semiring of
scalars
$\CC (\II  , \II )$. For \textbf{Rel} and real inner product spaces it is the
identity, while in the case of $\FdHilb$, it corresponds to \emph{complex conjugation}.

We need a compatibility condition between the strong compact closure and the biproducts.
\begin{definition}\em
We say that a category $\CC$ is a \em strongly compact closed category with biproducts \em iff
\ben
\item It is strongly compact closed;
\item It has biproducts;
\item The coproduct injections 
\[
q_i: A_i\to \bigoplus_{k=1}^{k=n}A_k
\]
satisfy 
\[
q_j^\dagger\circ q_i=\delta_{ij}\,.
\]
From this, it follows that  we can require that the chosen projections and
injections in Proposition~5.2 additionally satisfy $(p_i)^{\dagger}=q_i$.
\een
\end{definition}

\paragraph{Examples}
Finite-dimensional Hilbert spaces and real inner product spaces, categories of relations, and categories of free modules and semimodules are all examples of strongly compact closed categories with biproducts.

\subsection{Spectral Decompositions}

We define a \emph{spectral decomposition} of an object $A$ to be a unitary isomorphism 
\[
U : A \to \bigoplus_{i=1}^{i=n} A_i\,.
\]
(Here the `spectrum' is just the set of indices $1, \ldots , n$).
Given a spectral decomposition $U$, we define
morphisms 
\beqa
\neoiota_j &\!\!\!:=\!\!\!& U^{\dagger} \!\circ q_{j}: A_{j} \to A\\
\pi_j &\!\!\!:=\!\!\!&  \psi_j^{\dagger} = p_j \circ U : A \to A_{j}\,,
\eeqa
diagramatically  
\begin{diagram}
A_j & \rTo^{\neoiota_j} & A \\
\dTo^{q_j}& \ldTo_{U} & \dTo_{\pi_j}\\
\ \ \bigoplus_{i=1}^{i=n} A_i& \rTo_{p_j} & A_j
\end{diagram}
and finally \emph{projectors} 
\[
{\rm P}_j := \neoiota_j
\circ \pi_j : A \to A\,.
\] 
These projectors are \emph{self-adjoint}
\[ \PP_j^{\dagger} = (\neoiota_j \circ \pi_j )^{\dagger} =
\pi_j^{\dagger} \circ \neoiota_j^{\dagger} = \neoiota_j \circ \pi_j = \PP_j
\]
\emph{idempotent} and \emph{orthogonal}:
\[
{\rm P}_i\circ {\rm P}_j=\neoiota_i\circ \pi_i\circ \neoiota_j
\circ \pi_j=\neoiota_i\circ \delta_{ij}\circ \pi_j=  
\delta_{ij}^A\circ{\rm P}_i .
\]
Moreover, they yield a \emph{resolution of the identity}:
\beqa
\sum_{i=1}^{i=n} \PP_i &= & \sum_{i=1}^{i=n} \neoiota_i \circ\pi_i\\
&= &\sum_{i=1}^{i=n} U^{\dagger} \circ q_i \circ p_i \circ U \\ 
&= & U^{\dagger} \circ (\sum_{i=1}^{i=n} q_i \circ p_i )
\circ U\\ 
&= & U^{-1} \circ 1_{\bigoplus_{i}\! A_i} \circ
U = 1_{A} \, .
\eeqa

\subsection{Bases and dimension}

A \em basis \em for an object $A$ is a unitary isomorphism 
\[
{\sf base}:n\cdot\II\to A\,.
\]
Given bases ${\sf base}_A$ and ${\sf base}_B$ for objects $A$ and $B$ respectively we can define the matrix $(m_{ij})$ of any
morphism
$f:A\to B$ in those two bases as the matrix of 
\[
{\sf base}_B^\dagger\circ f\circ{\sf base}_A:n_A\cdot\II\to n_B\cdot\II
\]
as in Section \ref{sec:biprods}.

\begin{proposition}
Given $f:A\to B$ and 
\[
{\sf base}_A:n_A\cdot\II\to A
\quad{\rm and}\quad{\sf base}_B:n_B\cdot\II\to A
\] 
the matrix $(m'{\!\!}_{ij})$ of $f^\dagger$ in these bases is the conjugate transpose of the matrix
$(m_{ij})$ of
$f$.
\end{proposition}
\bpf
We have 
\beqa
m'{\!\!}_{ij}
&\!\!=\!\!&p_i\circ {\sf base}_A^{\dagger} \circ f^\dagger \circ{\sf base}_B \circ q_j\\
&\!\!=\!\!&(p_j\circ{\sf base}_B^\dagger\circ f\circ{\sf base}_A\circ q_i)^\dagger\\
&\!\!=\!\!& m_{ji}^\dagger\,.
\eeqa 
\vspace{-11.0mm} 

\hfill\endproof\newline

\noindent If in addition to the assumptions of Proposition \ref{pr:adjIn} and Proposition
\ref{pr:UnIn} there exist bases for $A$ and $B$, we can prove converses to both
of them.

\begin{proposition}\label{pro:inprod2}
If there exist bases for $A$ and $B$
then $f:A\to B$ is the adjoint to $g:B\to A$ if and only if
\[
\langle f\circ\psi\mid \phi\rangle_B=\langle \psi\mid g\circ\phi\rangle_A
\]
for all $\psi:\II\to A$ and $\phi:\II\to B$.
\end{proposition}
\bpf
Let $(m_{ij})$ be the matrix of $f^\dagger$ and $(m'{\!\!}_{ij})$ the matrix of $g$ in the given bases.
By Proposition \ref{pro:inprod} we have
\beqa
m_{ij}
\!\!\!&=&\!\!\!p_i\circ{\sf base}_A^\dagger\circ\! f^\dagger\circ {\sf base}_B\circ q_j\\
\!\!\!&=&\!\!\!\langle f\circ {\sf base}_A\!\circ q_i\mid {\sf base}_B\!\circ q_j\rangle_B\\ 
\!\!\!&=&\!\!\!\langle f\circ\psi\mid \phi\rangle_B=\langle \psi\mid g\circ\phi\rangle_A\\
\!\!\!&=&\!\!\!\langle  {\sf base}_A\!\circ q_i\mid g\!\circ{\sf base}_B\!\circ q_j\rangle_A\\
\!\!\!&=&\!\!\!p_i\circ{\sf base}_A^\dagger\circ\! g\circ {\sf base}_B\circ q_j=m'{\!\!}_{ij}\,.
\eeqa
Hence the matrix elements of $g$ and $f^\dagger$ coincide so $g$ and $f^\dagger$ are equal.
The converse is given by Proposition \ref{pr:adjIn}.
\hfill\endproof

\begin{proposition}\label{pr:UnIn2}
If there exist bases for $A$ and $B$ then a morphism $U:A\to B$ is unitary
if and only if it preserves the inner product, that is 
for all $\psi,\phi:\II\to A$ we have 
\[
\langle U\circ\psi\mid U\circ\phi\rangle_B=
\langle \psi\mid \phi\rangle_A\,.
\]
\end{proposition}
\bpf
We have 
\[
\langle U^{-1}\!\!\circ\psi\mid \phi\rangle_A=
\langle U\circ U^{-1}\!\!\circ\psi\mid U\circ \phi\rangle_B=
\langle \psi\mid U\circ \phi\rangle_B
\]
and hence by Proposition \ref{pro:inprod2}, $U^\dagger=U^{-1}$. The converse is
given by Proposition \ref{pr:UnIn}.
\hfill\endproof\newline

\noindent Note also that when a basis is available we can assign to $\psi^\dagger:A\to \II$ and $\phi:\II\to A$
matrices
\[
\left(
\begin{array}{ccc}
\psi_1^\dagger&
\cdots&
\psi_n^\dagger
\end{array}
\right)
\qquad\qquad
\left(
\begin{array}{c}
\phi_1\\
\vdots\\
\phi_n
\end{array}
\right)
\]
respectively, and by Proposition \ref{pro:inprod}, the
inner product becomes
\[
\langle \psi \mid \phi \rangle =\left(
\begin{array}{ccc}
\psi_1^\dagger&
\cdots&
\psi_n^\dagger
\end{array}
\right)
\left(
\begin{array}{c}
\phi_1\\
\vdots\\
\phi_n
\end{array}
\right)
=\sum_{i=1}^{i=n}\psi_i^\dagger\circ\phi_i\,. 
\]

Interestingly, two different notions of dimension arise in our setting.  We
assign an \em integer dimension \em ${\sf dim}(A)\in\mathbb{N}$ to an object
$A$ provided there exists a base 
\[
{\sf base}:{\sf dim}(A)\cdot\II\to A\,.
\] 
Alternatively, we introduce the \em scalar dimension \em as
\[
{\sf dim}_s(A):=\epsilon_A\circ\sigma_{A^*\!\!,A}\circ\eta_A\in{\bf C}(\II,\II) .
\]
We also have:
\[
{\sf dim}_s(\II)=1_\II
\qquad\quad
{\sf dim}_s(A^*)={\sf dim}_s(A)
\]
\[
{\sf dim}_s(A\otimes B)={\sf dim}_s(A){\sf dim}_s(B)
\]
In ${\bf FdVec}_\mathbb{K}$ these notions of dimension coincide, in the sense that ${\sf dim}_s(V)$ is multiplication with the
scalar
${\sf dim}(V)$. In ${\bf Rel}$ the integer dimension corresponds to the cardinality of the set, and is only well-defined for finite sets, while ${\sf
dim}_s(X)$ always exists;  however,  ${\sf
dim}_s(X)$ can only take two values, $0_\II$ and $1_\II$, and the two notions of dimension diverge for sets of cardinality greater than 1.

\Section{Abstract quantum mechanics}\label{sec:absquantprot}  

We can identify the basic ingredients of finitary quantum mechanics in any
strongly compact closed category with
biproducts.  
\bit
\item[{\bf 1.}] A \emph{state space} is represented by an object $A$.
\item[{\bf 2.}] A \em basic variable \em (`type of qubits') is a state space $Q$ with a 
given unitary isomorphism 
\[
{\sf base}_Q:{\rm I}\oplus{\rm I}\to Q
\]
which we call the
\em computational basis \em of $Q$. By using the isomorphism $n \cdot \II \simeq
(n\cdot\II )^*$ described in Section~5, we also obtain a computational basis for
$Q^*$.
\item[{\bf 3.}] A \emph{compound system} for which the subsystems are  described by
$A$ and $B$ respectively is
described by $A\otimes B$. If we have computational bases ${\sf base}_A$
and ${\sf base}_B$, then we define\vspace{-1mm} 
\[ 
{\sf base}_{A\otimes B}:=({\sf base}_A\otimes{\sf base}_B)\circ d_{nm}^{-1}
\] 
where 
\[
d_{nm} : n \cdot \II \otimes m \cdot \II \simeq (nm) \cdot \II
\]
is the canonical isomorphism constructed using first the left distributivity
isomorphism $\upsilon$, and then the right distributivity isomorphism
$\tau$, to give the usual lexicographically-ordered computational
basis for the tensor product.
\item[{\bf 4.}] Basic data transformations are unitary isomorphisms.
\item[{\bf 5a.}] A \emph{preparation} in a state space $A$ is a morphism 
\[
\neoiota:{\rm I}\to A
\] 
for which there exists a unitary $U:\II\oplus B \to A$ such that 
\begin{diagram}
{\rm I} & \rTo^{\neoiota} & A \\
\dTo^{q_1}& \ruTo_{U} & \\
{\rm I}\oplus B& 
\end{diagram}
commutes.
\item[{\bf 5b.}] Consider a spectral decomposition 
\[
U : A \to \bigoplus_{i=1}^{i=n} A_i
\]
with associated
projectors $\PP_j$. This gives rise to the \emph{non-destructive measurement}
\[ \langle\PP_i \rangle_{i=1}^{i=n}:A\to n\cdot A .
\]
The projectors 
\[
\PP_i:A\to A
\]
for $i=1, \ldots , n$ are called the \emph{measurement branches}.
This measurement is \emph{non-degenerate} if $A_i = \II$ for all $i = 1,
\ldots , n$. In this case we refer to $U$ itself as a \emph{destructive measurement} or  \emph{observation}. The  morphisms 
\[
\pi_i = p_i \circ
U :A\to\II
\]
for $i=1, \ldots , n$ are called \emph{observation branches}.
(We leave discussion of `degenerate destructive measurements', along with other variant notions of measurement, to future work).
\eit
Note that the type of a non-destructive measurement makes it explicit that it is an operation which
involves  an indeterministic transition (by contrast with the standard Hilbert space quantum
mechanical formalism). 
\bit
\item[{\bf 6a.}]  Explicit biproducts represent the \emph{branching} arising from the indeterminacy
of measurement outcomes. 
\eit
Hence an operation $f$ acting on an explicit biproduct $A\oplus B$ should itself be an explicit
biproduct, \textit{i.e.}~we want 
\[
f=f_1\oplus f_2:A\oplus B\to C\oplus D\,,
\]
 for $f_1:A\to C$ and $f_2:B\to D$.
The dependency of $f_i$ 
on the branch it is in captures \emph{local} classical communication.
The full force of non-local classical communication is enabled by Proposition \ref{distributivity}.
\bit
\item[{\bf 6b.}]  Distributivity isomorphisms represent \em non-local classical communication\em.
\eit
To see this, suppose e.g. that we have a compound system $Q \otimes A$, and we
(non-destructively) measure the qubit in the first component, obtaining a new system state 
described by
$(Q \oplus Q) \otimes A$. At this point, we know `locally',
\textit{i.e.} at the site of the first component, what the measurement 
outcome is, but we have not propagated this information to the rest
of the system $A$. However, after applying the distributivity
isomorphism
\[ (Q \oplus Q) \otimes A \simeq (Q \otimes A) \oplus (Q \otimes 
A) \]
the information about the outcome of the measurement on the first
qubit has been propagated globally throughout the system, and we can
perform operations on $A$ depending on the measurement outcome, e.g.
\[ (1_Q \otimes U_0 )  \oplus (1_Q \otimes U_1 )  \]
where $U_0$, $U_1$ are the  operations we wish to perform on $A$ in
the event that the outcome of the measurement we performed on $Q$ was
0 or 1 respectively.

\subsection*{The Born rule} 

We now show how the \emph{Born rule}, which is the key quantitative feature of
quantum mechanics, emerges automatically from our abstract setting.  

For a preparation $\neoiota : \II \to A$ and 
spectral decomposition 
\[
U : A \to \bigoplus_{i=1}^{i=n} A_i\,,
\] 
with
corresponding non-destructive measurement 
\[
\langle\PP_i \rangle_{i=1}^{i=n}:A\to n\cdot A\,,
\]
we can consider the protocol
\begin{diagram}
\II & \rTo^{\neoiota} & A & \rTo^{\langle\PP_i \rangle_{i=1}^{i=n}} & n \cdot
A\,.
\end{diagram}
We define scalars
\[ 
{\sf Prob}({\rm P}_i,\neoiota) := \langle \neoiota \mid \PP_i\mid \neoiota
\rangle =  
\neoiota^{\dagger} \circ \PP_i \circ \neoiota\,. 
\]

\begin{proposition}
With notation as above, 
\[
{\sf Prob}({\rm P}_i,\neoiota) = ({\sf Prob}({\rm
P}_i,\neoiota))^{\dagger}
\]
and
\[
\sum_{i=1}^{i=n}{\sf Prob}({\rm P}_i,\neoiota) = 1\,.
\]
Hence we think of the scalar ${\sf Prob}({\rm P}_j,\neoiota)$ as 
`the probability of obtaining  the $j$'th outcome of the measurement
$\langle\PP_i
\rangle_{i=1}^{i=n}$ on the state $\neoiota$'. 
\end{proposition}

\bpf 
From the definitions of preparation and the projectors, there are
unitaries $U$, $V$ such that 
\[
{\sf Prob}({\rm P}_i,\neoiota) = (V \circ q_{1})^{\dagger} \circ U^\dagger
\circ q_i \circ p_i \circ U
\circ V \circ q_1\]  for each $i$. Hence
\beqa
\sum_{i=1}^{i=n} {\sf Prob}({\rm P}_i,\neoiota)
\!\!\!\!&= &\!\!\!\!\sum_{i=1}^{i=n}
p_1
\circ V^\dagger \circ U^\dagger
\circ q_i \circ p_i \circ U \circ V \circ q_1 \\  
\!\!\!\!&= &\!\!\!\!p_1 \circ 
V^\dagger
\!\circ U^\dagger \!\circ\! \Bigl(\sum_{i=1}^n \!q_i
\circ  p_i \Bigr)\! \circ
 U \circ V \circ q_1 \\
\!\!\!\!&= &\!\!\!\! p_1 \circ  V^{-1}\! \circ U^{-1}\! \circ 1_{n\cdot\II} \circ U 
\circ V
\circ q_1 \\  \!\!\!\!&= &\!\!\!\! p_1 \circ q_1 = 1_{\II}\,.\mbox{\hspace{3.43cm}\endproof}
\eeqa 
\vspace{1mm}
Moreover, since by definition
${\rm
P}_j=\pi_j^\dagger\circ\pi_j$,
we can rewrite the Born rule expression as 
\beqa
{\sf Prob}({\rm P}_j,\neoiota)
\!\!\!&= &\!\!\!\neoiota^{\dagger} \circ \PP_j\circ \neoiota\\
\!\!\!&= &\!\!\!\neoiota^{\dagger} \circ \pi_j^\dagger\circ\pi_j\circ \neoiota\\
\!\!\!&= &\!\!\!(\pi_j\circ\neoiota)^\dagger\circ\pi_j\circ \neoiota\\
\!\!\!&= &\!\!\!s_j^\dagger\circ s_j 
\eeqa
for some scalar $s_j\in{\bf C}(\II,\II)$.  Thus $s_j$ can be thought of as the `probability amplitude' giving rise to the probability $s_j^\dagger\circ s_j$, which is of course self-adjoint. If we consider the protocol 
\begin{diagram}
\II & \rTo^{\neoiota} & A & \rTo^{\langle \pi_i \rangle_{i=1}^{i=n}} & n
\cdot \II\,.
\end{diagram}
which involves an observation $\langle \pi_i \rangle_{i = 1}^{i =n}$, then
these scalars $s_j$ correspond to the branches 
\begin{diagram}
\II & \rTo^{\neoiota} & A & \rTo^{\pi_j } & \II\,.
\end{diagram}

\Section{Abstract quantum protocols}

\noindent We prove correctness of the example protocols.

\subsection{Quantum teleportation} 

\begin{definition}\em
A \em teleportation base \em is a scalar $s$ together with a  
morphism 
\[
{\sf prebase}_{\rm T} : 4\cdot{\rm I}\to Q^*\otimes Q
\]
such that:
\begin{itemize}
\item ${\sf base}_{\rm T} := s \sdot {\sf prebase}_{\rm T}\,$ is unitary.
\item the four maps $\beta_j:Q\to Q$, where $\beta_j$ is defined by 
\[
\uu\beta_j\uuu:={\sf prebase}_{\rm
T}\circ q_j\,,
\]
are unitary. 
\item $2 s^{\dagger} s = 1$.
\end{itemize}
The morphisms $s \sdot \uu\beta_j\uuu$ are the \em base 
vectors \em of the teleportation base.   A teleportation base is a  
\em Bell base \em when
the \em 
Bell base  maps \em 
\[
\beta_1,\beta_2,\beta_3,\beta_4:Q\to Q
\]
satisfy\footnote{This choice of axioms is sufficient for our purposes.  One might prefer to axiomatize a notion of Bell base such that the corresponding Bell base maps are exactly the Pauli matrices.}
\[
\beta_1=1_Q\quad \beta_2=\sigma_{Q}^{\oplus}\quad \beta_3=\beta^\dagger_3\quad
\beta_4=\sigma_{Q}^{\oplus}\circ\beta_3
\]
where 
\[
\sigma_{Q}^{\oplus}:={\sf base}_Q^{}\circ\sigma^\oplus_{\II,\II}\circ{\sf
base}_Q^{-1}\,.
\]
A teleportation base defines a \em teleportation observation \em
\[
\langle s^{\dagger} \sdot 
\dd
\beta_i\ddd\rangle_{i=1}^{i=4}:Q\otimes Q^*\to 4\cdot{\rm
I}\,.
\]
\end{definition}

To emphasize the
identity of the individual qubits we label the three copies of $Q$ we
shall consider as $Q_a$, $Q_b$, $Q_c$. We also use labelled
identities,
e.g.~${1_{bc} : Q_b \rightarrow Q_c}$, and labelled Bell bases.
Finally, we introduce 
\[
\Delta_{ac}^4:=\langle s^{\dagger}s \sdot 1_{ac}\rangle_{i=1}^{i=4}:
Q_a\to 4\cdot Q_c
\]
as the \emph{labelled,
weighted diagonal}. This expresses the intended behaviour of teleportation, namely that the
input qubit is propagated to the output along each branch of the
protocol, with `weight' $s^{\dagger}s$,  corresponding to the probability
amplitude for that branch. Note that the sum of the corresponding
probabilities is 
\[
4(s^{\dagger}s)^{\dagger}s^{\dagger}s = (2s^{\dagger} s)(2 s^{\dagger} s) =
1\,.
\]

\begin{theorem}\label{thm:teleport}
The following diagram commutes.
\begin{diagram}
Q_a&\rIs&Q_a\\ 
&&\dTo^{\rho_a}&\hspace{-1.5cm}{\bf import\ unknown\ state}\\
&&Q_a\otimes{\rm I}\\
&&\dTo^{1_a\otimes (s \sdot \uu 1_{bc}\!\!\uuu)}&\hspace{-1.5cm}{\bf produce\ EPR\mbox{\bf
-}pair}\\ &&Q_a\otimes(Q_b^*\!\otimes Q_c)\\
&&\dTo^{\alpha_{a,b,c}}&\hspace{-1.5cm}{\bf spatial\ delocation}\\
\dTo^{\Delta^4_{ac}}&&(Q_a\otimes Q^*_b)\otimes Q_c\\
&&\dTo^{\quad\quad\langle s^{\dagger} \sdot \dd
\beta_i^{ab}\!\!\ddd\rangle_{i=1}^{i=4}\otimes\! 1_c}&\hspace{-1.5cm}{\bf
teleportation\ observation}\\ &&\left(4\cdot{\rm
I}\right)\otimes Q_c\\
&&\dTo^{\left(4\cdot\lambda^{-1}_c\right)\!\circ\upsilon_c}&\hspace{-1.5cm}{\bf
classical\ communication}\\ &&\ \ \ 4\cdot Q_c\\
&&\dTo^{\bigoplus_{i=1}^{i=4}(\beta_i^c)^{-1}}&\hspace{-1.5cm}{\bf unitary\
correction}\\
\ \ \ \ 4\cdot Q_c&\rIs&4\cdot Q_c\!\!\!\!\!\! 
\end{diagram}
The right-hand-side of the above diagram is our formal description of
the teleportation protocol; the commutativity of the diagram expresses 
the correctness of the protocol.
Hence any strongly compact closed category with biproducts admits  
quantum teleportation provided it contains a teleportation base.
If we do a Bell-base observation then the corresponding
unitary corrections are
\[
\beta^{-1}_i\!\!=\beta_i\ \ {\it for}\ \ i\in\{1,2,3\}
\quad{\it and}\quad \beta_4^{-1}\!\!=\beta_3\circ\sigma_{Q}^{\oplus}\,.
\]
\end{theorem}
\bpf
For a proof of the commutativity of this diagram see the
Appendix -- it uses the universal property of the
product, Lemma
\ref{lm:compos}, naturality of $\lambda$ and the explicit form of

\[
\upsilon_c:=\langle p_i^{\rm I}\!\otimes\! 1_c\rangle_{i=1}^{i=4}\,.
\]
In the
specific case of a Bell-base observation we use $1_Q^\dagger=1_Q\,,\quad (\sigma_{Q}^{\oplus})^\dagger=\sigma_{Q}^{\oplus}$ and $(\sigma_{Q}^{\oplus}\circ\beta_3)^\dagger=
\beta_3^\dagger\circ(\sigma_{Q}^{\oplus})^\dagger=
\beta_3\circ\sigma_{Q}^{\oplus}$.
\hfill\endproof\newline

Although in {\bf Rel} teleportation works for `individual observational branches' it fails to
admit the full teleportation protocol since there are only two
automorphisms of $Q$ (which is just a two-element set, \ie the type of 
`classical bits'), and hence
there is no teleportation base.

We now consider sufficient conditions on the ambient category $\CC$
for a teleportation base to exist.
We remark firstly that if ${\bf C}({\rm I},{\rm I})$ contains an
additive inverse for $1$, then it is a ring, and moreover all additive 
inverses exist in each hom-set $\CC (A, B)$, so $\CC$ is enriched over 
Abelian groups.
Suppose then that  ${\bf C}({\rm I},{\rm I})$ is a ring with $1\not= -1$.
We can define a morphism 
\[
{\sf prebase}_{\rm T}= {\sf base}_{Q^*\otimes Q} \circ M :4\cdot{\rm I}\to
Q^*\!\otimes Q
\]
where $M$ is
the endomorphism of $4 \cdot \II$ determined by the matrix
\[
\left(
\begin{array}{cccc}
1&0&1&0\\
0&1&0&1\\
0&1&0&\!\!\!\!-\!1\\
1&0&\!\!\!\!-\!1&0
\end{array}
\right)
\]
The corresponding morphisms $\beta_j$ will have $2 \times 2$ matrices determined
by the columns of this $4 \times 4$ matrix, and will be unitary.
If ${\bf C}({\rm I},{\rm I})$ furthermore contains a scalar $s$
satisfying $2s^{\dagger} s = 1$, then $s \sdot {\sf
prebase}_{\rm T}$ is unitary, and the conditions for a teleportation base are fulfilled.
Suppose we start with a ring $R$ containing an element $s$ satisfying
$2s^2 = 1$. (Examples are plentiful, e.g. any  subring of $\mathbb{C}$, or of
$\mathbb{Q}(\sqrt{2})$, containing $\frac{1}{\sqrt{2}}$). The category 
of finitely generated free $R$-modules and $R$-linear maps is strongly 
compact
closed with biproducts, and admits a teleportation base (in which $s$
will appear as a scalar with $s = s^{\dagger}$), hence
realizes teleportation.

\subsection{Logic-gate teleportation}

Logic gate teleportation of qubits requires only a minor modification as
compared to the teleportation protocol.
\begin{theorem}\label{thm:logicgate}
Let unitary morphism $f:Q\to Q$ be such that for each $i\in\{1,2,3,4\}$ a
morphism
$\varphi_i(f):Q\to Q$ satisfying 
\[ 
f\circ\beta_i=\varphi_i(f)\circ f
\]
exists.
The diagram of Theorem \ref{thm:teleport} with the modifications made
below commutes.   
\begin{diagram}
\dDots&&\dDotsto\\
&&Q_a\otimes{\rm I}\\
&&\dTo^{1_a\otimes(s \sdot \uu f\uuu)}&\hspace{-5mm}{\bf produce}\
f{\bf\mbox{\bf -}state}\\  &&Q_a\otimes(Q^*_b\!\otimes
Q_c)\\ 
\dTo^{\Delta^4_{ac}\!\circ\! f}&&\dDotsto\\
&&\ \ \ 4\cdot Q_c\\
&&\dTo^{\bigoplus_{i=1}^{i=4}(\varphi_i(f))^{-1}}&\hspace{-5mm}{\bf unitary\ correction}\\ 
\ \ \ \ 4\cdot Q_c&\rIs&4\cdot Q_c\!\!\!\!\!\! 
\end{diagram}
The right-hand-side of the diagram is our formal description of
logic-gate teleportation of $f:Q\to Q$; the commutativity of the diagram under
the stated conditions expresses the correctness of logic-gate teleportation for qubits.
\end{theorem}
\bpf
See the diagram in the appendix. 
\hfill\endproof\newline

This two-dimensional case does not yet provide a universal
computational primitive, which requires teleportation of $Q\otimes
Q$-gates \cite{Gottesman}. We present the example of teleportation of a
$\cnot$ gate \cite{Gottesman} (see also \cite{Coe1} Section 3.3).  

Given a Bell base we define a $\cnot$ gate as one
which acts as follows on tensors of the Bell base maps\footnote{One could give a more
explicit definition of a $\cnot$ gate, e.g.~by specifying the matrix.  However, our
generalized definition suffices to provide the required corrections. Moreover, this
example nicely illustrates the attitude of `focussing on the essentials by abstracting'.}:
\beqa
\cnot\circ (\sigma_{Q}^{\oplus}\otimes 1_Q)&\!\!=\!\!&(\sigma_{Q}^{\oplus}\otimes
\sigma_{Q}^{\oplus})\circ\cnot\\
\cnot\circ (1_Q\otimes \sigma_{Q}^{\oplus})&\!\!=\!\!&(1_Q\otimes
\sigma_{Q}^{\oplus})\circ\cnot\\ 
\cnot\circ (\beta_3\otimes 1_Q)&\!\!=\!\!&(\beta_3\otimes 1_Q)\circ\cnot\\
\cnot\circ (1_Q\otimes \beta_3)&\!\!=\!\!&(\beta_3\otimes \beta_3)\circ\cnot
\eeqa
It follows from this that  
\beqa
\cnot\circ (\beta_4\otimes 1_Q)&\!\!=\!\!&(\beta_4\otimes
\sigma_{Q}^{\oplus})\circ\cnot\\
\cnot\circ (1_Q\otimes \beta_4)&\!\!=\!\!&(\beta_3\otimes
\beta_4)\circ\cnot
\eeqa
from which in turn it follows by bifunctoriality of the tensor that the required unitary
corrections factor into single qubit actions, for which we introduce a notation by setting
\beqa
\cnot\circ (\beta_i\otimes 1_Q)&\!\!=\!\!&\varphi_1(\beta_{i})\circ\cnot\\
\cnot\circ (1_Q\otimes \beta_i)&\!\!=\!\!&\varphi_2(\beta_{i})\circ\cnot
\eeqa
The reader can verify that for
\[
4^2\cdot
(Q_{c_1}\!\!\otimes\! Q_{c_2}):=4\cdot (4\cdot (Q_{c_1}\!\!\otimes\! Q_{c_2}))
\]
and 
\[
\hspace{-1.2mm}\Delta_{ac}^{4^2}\!:=\!\langle s^{\dagger}\!s \sdot \langle s^{\dagger}\!s
\sdot\! 1_{ac}\rangle_{i=1}^{i=4}\rangle_{i=1}^{i=4}\!:\!
Q_{a_1}\!\!\otimes\! Q_{a_2}\!\to 4^2\!\cdot 
(Q_{c_1}\!\!\otimes\! Q_{c_2})\hspace{-1.2mm}
\] 
the following diagram commutes.

\begin{diagram}
Q_{a_1}\!\!\otimes\! Q_{a_2}&\rIs&Q_{a_1}\!\!\otimes\! Q_{a_2}\\   %
&&\dTo^{\rho_a}&\hspace{-1.7cm}{\bf import\ unknown\ state}\\  %
&&(Q_{a_1}\!\!\otimes\! Q_{a_2})\otimes{\rm I}\\ %
&&\dTo^{\hspace{-1cm}1_a\otimes (s^2 \sdot \uu \cnot\uuu)}&\hspace{-1.7cm}{\bf produce\
\cnot\mbox{\bf -}state}\\ &&\hspace{-2cm}(Q_{a_1}\!\!\otimes\!
Q_{a_2})\otimes((Q_{b_1}\!\!\otimes\! Q_{b_2})^*\!\otimes (Q_{c_1}\!\!\otimes\!
Q_{c_2}))\hspace{-2cm}\\
&&\dTo^{\hspace{-2cm}(\alpha,\sigma)\!\circ(1_a\!\otimes(u_{b}\!\otimes\!
1_c))}&\hspace{-1.7cm}{\bf   spatial\ delocation}\\ &&\hspace{-2.3cm}((Q_{a_1}\!\!\otimes\!
Q_{b_1}^*)\otimes (Q_{c_1}\!\!\otimes\! Q_{c_2}))\otimes(Q_{a_2}\!\!\otimes\!
Q_{b_2}^*)\hspace{-2.3cm}\\ &&\dTo^{\hspace{-2cm}(\langle s^{\dagger} \!\!\sdot\!
\dd
\beta_i^{a_1\!b_1\!}\!\!\ddd\rangle_{i=1}^{i=4}\!\otimes\!\! 1_{c})\!\otimes\! 
\!1_{2}}&\hspace{-1.7cm}{\bf 1st\ observation}\\ &&\hspace{-2.3cm}(\left(4\cdot{\rm
I}\right)\otimes (Q_{c_1}\!\!\otimes\!Q_{c_2}))\otimes (Q_{a_2}\!\!\otimes\!
Q_{b_2}^*)\hspace{-2.3cm}\\
\dTo~{\hspace{-6mm}\Delta^{4^2}_{ac}\circ\cnot}&&\dTo^{((4\cdot\lambda^{-1}_c)\!\circ\!\upsilon_c)\otimes
1_2}&\hspace{-1.7cm}{\bf 1st\ communication}\\ 
&&\hspace{-2.3cm}(4\cdot(Q_{c_1}\!\!\otimes\!Q_{c_2}))\otimes
(Q_{a_2}\!\!\otimes\! Q_{b_2}^*)\hspace{-2.3cm}\\
&&\dTo^{\hspace{-2.3cm}\left(\bigoplus_{i=1}^{i=4}(\varphi_1^c(\beta_{i}))^{-1}\right)\!\otimes
\! 1_2}&\hspace{-1.7cm}{\bf 1st\ correction}\\
&&\hspace{-2.3cm}(4\cdot(Q_{c_1}\!\!\otimes\! Q_{c_2}))\otimes
(Q_{a_2}\!\!\otimes\! Q_{b_2}^*)\hspace{-2.3cm}\\
&&\dTo^{\hspace{-2cm}(4\cdot 1_c)\!\otimes\!\langle s^{\dagger} \!\sdot\!
\dd
\beta_i^{a_2\!b_2\!}\!\!\ddd\rangle_{i=1}^{i=4}}
&\hspace{-1.7cm}{\bf 2nd\ observation}\\
&&(4\cdot(Q_{c_1}\!\!\otimes\!Q_{c_2}))\otimes (4\cdot{\rm
I})\\ 
&&\dTo^{(4\cdot\rho^{-1}_{4c})\!\circ\!\tau_{4c}}&\hspace{-1.7cm}{\bf
2nd\ communication}\\ 
&&(4\cdot(4\cdot(Q_{c_1}\!\!\otimes\!Q_{c_2})))\\
&&\dTo^{\hspace{-2cm}\bigoplus_{i=1}^{i=4}(4\cdot
\varphi_2^c(\beta_{i}))^{-1}}&\hspace{-1.7cm}{\bf 2nd\ correction}\\
\ \ \ \ 4^2\cdot
(Q_{c_1}\!\!\otimes\! Q_{c_2})&\rIs&4^2\cdot
(Q_{c_1}\!\!\otimes\! Q_{c_2})\!\!\!\!\!\! 
\end{diagram}

\subsection{Entanglement swapping}

\begin{theorem}\label{thm:swap}
Setting\hspace{-2.5mm}
\[
\begin{array}{lcl}
\gamma_i&\!\!:=\!\!&(\beta_i)_*\vspace{1.5mm}\\
{\rm
P}_{i}&\!\!:=\!\!&s^{\dagger}s\sdot(\uu\gamma_i\uuu\circ\dd\beta_i\ddd)\vspace{1.5mm}\\ 
\zeta_i^{ac\!\!}&\!\!:=\!\!&\bigoplus_{i=1}^{i=4}\left((1_b^*\otimes
\gamma_i^{-1})\otimes(1_d^*\otimes\beta_i^{-1})\right)\vspace{1.5mm}\\ 
\Theta_{ab\!\!}&\!\!:=\!\!&1_d^*\otimes\langle {\rm
P}_{i}\rangle_{i=1}^{i=4}\otimes\! 1_c\vspace{1.5mm}\\
\Omega_{ab\!\!}&\!\!:=\!\!&\langle s^{\dagger}s^3\sdot(\uu 1_{ba}\!\uuu\!\otimes\! \uu
1_{dc}\!\uuu)\rangle_{i=1}^{i=4}
\end{array}
\]   

\hspace{-2.5mm}\noindent
the following diagram commutes.  
\begin{diagram}
{\rm I}\otimes{\rm I}&\rIs&{\rm I}\otimes{\rm I}\\ 
&&\dTo^{s^2\sdot(\uu 1_{da}\!\uuu\!\otimes \!\uu 1_{bc}\!\uuu)}&\hspace{-1.8cm}{\bf produce\ 
EPR\mbox{\bf -}pairs}\\ &&(Q_d^*\otimes Q_a)\otimes(Q_b^*\!\otimes Q_c)\\ 
&&\dTo^{\alpha}&\hspace{-1.8cm}{\bf spatial\ delocation}\\
&&Q_d^*\otimes (Q_a\otimes Q^*_b)\otimes Q_c\\
\dTo^{\Omega_{ab\!\!}}&&\dTo^{\Theta_{ab}}&\hspace{-1.8cm}{\bf
Bell\mbox{\bf -}base\ measurement}\\ &&\hspace{-3mm}Q_d^*\otimes \left(4\cdot (Q_a\!\otimes
Q^*_b)\right)\otimes Q_c\\
&&\dTo^{(4\cdot(\alpha,\sigma))\circ(\tau,\upsilon)}&\hspace{-1.8cm}{\bf
classical\ communication}\\ &&\hspace{-3mm}4\cdot (\left(Q^*_b\!\otimes Q_a
\right)\!\otimes\!(Q_d^*\otimes Q_c))\\
&&\dTo^{\hspace{-3cm}
\zeta_i^{ac}} &\hspace{-1.8cm}{\bf unitary\  
correction}\\
4\cdot (\left(Q^*_b\!\otimes Q_a
\right)\hspace{-1.15cm}&&\hspace{-1cm}\hspace{-1.0mm}\otimes(Q_d^*\otimes Q_c))\hspace{3mm}
\end{diagram}
The right-hand-side of the above diagram is our formal description of 
the entanglement swapping protocol.
\end{theorem}
\bpf
See the diagram in the appendix --- it uses Lemma \ref{lm:precompos} and Lemma \ref{lm:CUT}. 
\hfill\endproof\newline 

We use $\gamma_i=(\beta_i)_*$ rather than $\beta_i$ to make ${\rm
 P}_i$ an endomorphism and hence a projector. 
The general definition of a `bipartite entanglement projector' is
\[
\hspace{-1.5mm}{\rm P}_f:=\uu f\uuu\circ\dd f_*\ddd=\uu
f\uuu\circ\dd f^\dagger\!\!\ddd\circ \sigma_{A^*\!, B}:A^*\otimes B\to A^*\otimes
B\hspace{-1.5mm}
\]
for $f:A\to B$, so in fact ${\rm P}_i={\rm P}_{(\beta_i)_*}$.

\Section{Conclusion}  

\paragraph{Other work.}
Birkhoff and von Neumann
\cite{BvN} attempted to capture quantum behavior abstractly in
lattice-theoretic terms --- see also Mackey \cite{Mackey} and
Piron \cite{Piron}.   The weak spot of this programme was 
the lack of a satisfactory treatment of compound systems --- whereas in our
approach the tensor $\otimes$ is a primitive.  Different kinds
of lattices do arise naturally in our setting, but we leave a discussion
of this to future work.

Isham and Butterfield \cite{IshBut} have reformulated the
Kochen-Specker theorem in a topos-theoretic setting.  On the one hand, 
assuming that the tensor in a compact closed category is the
categorical product leads to triviality---the category is then
necessarily equivalent to the one-object one-arrow category---and in
this sense the compact closed and topos axioms are not compatible.  On the other hand, each
topos yields a strongly compact closed category with biproducts
as its category of relations.

The recent papers \cite{Sel,vTon} use categorical methods for giving 
semantics to a quantum programming language, and a quantum lambda
calculus, respectively. In both cases, the objectives, approach and
results are very different to those of the present paper. A more
detailed comparison must again be left to future work.

\paragraph{Further Directions.} This work has many possible lines for further
development. We mention just a few.

\begin{itemize}
\item Our setting seems a natural one for developing type systems to
  control quantum behaviour.

\item In order to handle protocols and quantum computations more
  systematically, it would be desirable to have an effective syntax, 
  whose design should be guided by the categorical semantics.

\item The information flow level of analysis using only the
  compact-closed structure allows some very elegant and convenient
  `qualitative' reasoning, while adding biproducts allows very
  fine-grained modelling and analysis. The interplay between these two 
  levels merits further investigation.

\item We have not considered mixed states and non-projective measurements in this paper, but
they
  can certainly be incorporated in our framework.

\item In this paper, we have only studied finitary Quantum Mechanics.
A significant step towards the infinite dimensional case is provided
by the previous work on
\em nuclear ideals in tensored
$*$-categories \em \cite{ABP}.  The `compactness' axiom for nuclear ideals (see
Definition 5.7 in \cite{ABP}) corresponds to
our Compositionality Lemma~3.4. One of the main intended models of
nuclear ideals is given by the category of all Hilbert spaces and
bounded linear maps.

\item Another class of compact closed categories with biproducts are
  the Interaction Categories introduced by one of the authors
  \cite{Abramsky}. One can consider linear-algebraic versions of
  Interaction Categories --- `matrices extended in time' rather than
  `relations extended in time' as in \cite{Abramsky}. Does this lead to a 
  useful notion of quantum concurrent processes?

\end{itemize}

\section*{Acknowledgements}  

Rick Blute and Prakash Panangaden suggested some improvements to an
earlier version of this paper.

\bibliographystyle{latex8}  
\bibliography{arXivqcc}    

\onecolumn    

\appendix 

\section*{Appendix: Diagramatic proofs}\label{sec:diagrams}       
\ \vspace{-0.8cm} 

\ \par\bigskip
\noindent{\bf Proof of Lemma \ref{lm:compos} (compositionality).} The top trapezoid
is the statement of  the Lemma. 
\vspace{0.5cm}
\begin{diagram}
A&&&&&&\rTo^{g\circ f}&&&&&&C\\ 
&\rdTo^{\rho_A}&&&&&&&&&&\ruTo^{\lambda^{-1}_C}&\\ 
&&A\otimes {\rm I}&&\rTo^{1_A\otimes\uu g\uuu}&&A\otimes B^*\!\otimes C&&\rTo^{\dd f\ddd\otimes 1_C}&&{\rm I}\otimes C&&\\ 
&&&\rdTo~{1_A\otimes\eta_B}&&\ruTo~{1_{A\otimes B^*}\!\!\otimes g}&&\rdTo~{f\otimes 1_{B^*\!\otimes C}}&&\ruTo~{\ \epsilon_B\otimes 1_C}&&&\\ 
\dTo^f&&\dTo^{f\otimes 1_{\rm I}}&&A\otimes B^*\!\otimes B&&&&B\otimes B^*\!\otimes C&&\uTo_{1_{\rm I}\otimes g}&&\uTo_g\\
&&&&&\rdTo_{\ \ f\otimes 1_{B^*\!\otimes B}\hspace{-2mm}}&&\ruTo_{\hspace{-1mm}1_{B^*\!\otimes B}\otimes g}&&&&&\\
&&B\otimes{\rm I}&&\rTo_{1_B\otimes\eta_B}&&B\otimes B^*\otimes B&&\rTo_{\epsilon_B\otimes 1_B}&&{\rm I}\otimes B&&\\
&\ruTo_{\rho_B}&&&&&&&&&&\rdTo_{\lambda^{-1}_B}&\\
B&&&&&&\rIs_{}&&&&&&B
\end{diagram}

\vspace{-8.4cm}\hspace{7.5cm}{\bf Lemma \ref{lm:compos}}

\vspace{6.0cm}\hspace{6.8cm}{\bf Compact closedness}
 
\vspace{2.5cm}\noindent
{\bf Proof of Lemma \ref{lm:CUT} (compositional CUT).} The top trapezoid
is the statement of  the Lemma.
\vspace{0.5cm}
\begin{diagram}
{\rm I}&&&&\rTo^{\uu h\circ g\circ f\uuu}&&&&A^*\!\!\otimes\! D\\ 
&\rdTo^{\rho_{\rm I}}&&&&&&\ruTo^{\rho^{-1}_A\!\otimes\!1_{D^*}}_{1_A\!\otimes\!
\lambda^{-1}_{D^*}}\ruTo(2,4)_{1_{A^*}\!\!\otimes\!(h\!\circ\! g)\qquad\quad}&\\  
&&{\rm I}\otimes {\rm I}&\rTo^{{\uu f\uuu}\!\otimes\!{\uu
h\uuu}}&A^*\!\!\otimes\! B\!\otimes\! C^*\!\!\otimes\! D&\rTo^{1_{A^*}\!\otimes\!
\dd g\ddd\!\otimes\! 1_D}&A^*\!\!\otimes\! {\rm I}\!\otimes\! D&&\\ 
&&\dTo^{\eta_A\!\otimes\! 1_{\rm I}}&\rdTo~{{\uu f\uuu}\!\otimes\! 1_{\rm I}}&\uTo~{1_{A^*}\!\!\otimes\!{\uu h\uuu}}&&&&\\ 
\dTo^{\ \ \ \ \eta_A}&&A^*\!\!\otimes\! A\otimes\!{\rm I}&\rTo^{1_{A^*}\!\!\otimes\! f\!\otimes\! 1_{\rm I}}&A^*\!\!\otimes\! B\!\otimes\!{\rm I}&\lTo^{\rho_{A^*\otimes B}}&A^*\!\!\otimes\! B&&\uTo~{1_{A^*}\!\!\otimes\!(h\!\circ\! g\!\circ\! f)}\\ 
&\ruTo_{\rho_{A^*\!\otimes A}}&&&&&&\luTo_{1_{A^*}\!\!\otimes\! f}&\\ A^*\!\!\otimes\!
A&\rIs &&&&&&&A^*\!\!\otimes\! A
\end{diagram} 

\vspace{-3.84cm}\hspace{9.97cm}{\bf Lemma \ref{lm:compos}} 

\vspace{-2.44cm}\hspace{7.5cm}{\bf Lemma \ref{lm:CUT}}

\newpage 

\noindent{\bf Proof of Lemma \ref{lm:expldaggaer} (adjoints to points).} The top
trapezoid is the statement of  the Lemma.  The cell labelled
\textbf{SMC} commutes by symmetric monoidal coherence.
\vspace{0.5cm}
\begin{diagram}
A&&&&&\rTo^{\qquad\psi^\dagger}&&&&&\II\\ 
\dTo^{\lambda_A}&\rdTo^{\rho_A}&&&&&&&&\ruTo^{\epsilon_A}&\dIs\\ 
\II\otimes A&\rTo^{\sigma_{\II,A}}&A\otimes\II&\rTo^{\!\!\!\!1_A\otimes
u_\II\!\!\!\!}&A\otimes\II^*&\rIs&A\otimes\II^*&\rTo^{1_A\otimes \psi_*}&A\otimes
A^*&&\\ 
\dIs&\rdTo~{\eta_{\II}\otimes 1_A}&&\rdTo~{1_A\otimes\eta_{\II}}&\uTo_{1_A\otimes\rho_\II}&& \uTo_{\sigma_{\II^*\!\!,A}}&&\uTo_{\sigma_{A^*\!\!,A}}\\ 
\II\otimes A&&(\II^*\!\!\otimes\II) \otimes A&\rTo_{\!\!\!\!\sigma_{\II^*\!\otimes\II,A}\!\!\!\!}&A\otimes (\II^*\!\!\otimes\II)&&\II^*\!\!\otimes A&\rTo^{\psi_*\otimes 1_A}&A^*\!\!\otimes
A&\rTo^{\epsilon_{A^*}}&\II\\ 
&\rdTo_{\eta_{\II^*}\!\otimes
1_A\!\!}&\dTo_{\sigma_{\II^*\!\! ,\II}\otimes
1_A}&&&\hspace{16mm}\ruTo_{\!\!\!\lambda^{-1}_{\II^*\!\otimes
A}}&&\ruTo_{\!\!\!\lambda^{-1}_{A^*\!\otimes
A}}&&\ruTo_{\!\!\!\!\rho^{-1}_\II\!\!=\lambda^{-1}_\II}\\
&&(\II\otimes\II^* )\!\otimes
A&\rTo_{\!\!\!\!\alpha^{-1}_{\II ,
\II^*\!\! , A}\!\!\!\!}&\II\otimes(\II^*\!\otimes
A)&\rTo_{1_\II\otimes (\psi_*\otimes 1_A )}&\II\otimes
A^*\!\otimes
A&\rTo_{1_\II\otimes\epsilon_{A^*}}&\II\otimes\II 
\end{diagram}
 
\vspace{-2.64cm}\hspace{2.1cm}{\bf Prop.~\ref{prop:triqcc}}
\hspace{4cm}{\bf SMC}

\vspace{-4cm}\hspace{8.05cm}{\bf Lemma \ref{lm:expldaggaer}} 

\vspace{0.95cm}\hspace{14.15cm}{\bf Prop.~\ref{prop:triqcc}}
 
\vspace{4.9cm}\noindent
\noindent{\bf Proof of Theorem \ref{thm:teleport} (quantum teleportation).} For each $j\in\{1,2,3,
4\}$ we have a diagram of the form below. The top trapezoid
is the statement of  the Theorem. We ignore  the scalars  -- which cancel out against each
other -- in this proof.
\vspace{0.5cm}
\begin{diagram}
\hspace{-1mm}Q_a&&&&&&\rTo^{\ \ \langle 1_{ac}\rangle_{i=1}^{i=4}\ \
\!\!\!}&&&&&&4\cdot Q_c\\
&\rdTo^{\,\rho_a}&&&&&&&&&&\!\!\!\!\!\ruTo^{\bigoplus_{i=1}^{i=4}(\beta_i^c)_i^{-1}\!\!}&\\
&&\hspace{-2mm}Q_a\!\otimes\!{\rm I}&\rTo^{\!\!1_a\!\otimes\!\uu 1_{bc}\uuu\!}&\!Q_a\!\otimes\!
Q^*_b\!\!\otimes\! Q_c\!&\rTo^{\!\langle\dd
\beta^{ab\!\!}_i\ddd\rangle_{i=1}^{i=4}\!\otimes\!\!
1_c\!\!}&\left(4\cdot{\rm 
I}\right)\otimes Q_c&\rTo^{\!\langle p_i^{\rm I}\!\otimes\!\! 
1_c\rangle_{i=1}^{i=4}}&
4\cdot\left({\rm I}\!\otimes\!
Q_c\right)&\rTo^{\!4\cdot\lambda^{-1}_c\!}&4\cdot Q_c\hspace{-2mm}&&\\
\dIs&&&&&\hspace{-6.5mm}\rdTo_{\dd \beta^{ab}_j\!\ddd\!\otimes\!
1_c\hspace{-2mm}}&\dTo_{p^{\rm I}_j\otimes\! 1_c}&&\dTo_{p^{{\rm I}\otimes
Q_c}_j}&&\dTo_{p_j^{Q_c}}&&\dTo_{p_j^{Q_c}}\\ &&Q_c&&\pile{\lTo^{\lambda^{-1}_c}\\
\rTo_{\lambda_c}}&&{\rm I}\otimes Q_c&\rTo_{1_{{\rm I}\otimes Q_c}}&{\rm I}\otimes
Q_c&\rTo_{\lambda^{-1}_c}&Q_c&&\\ &\ruTo_{\!\beta^{ac}_j}&&&&&&&&&&\rdTo_{(\beta_j^c)^{-1}\!}\\
\hspace{-1mm}Q_a&&&&&&\rTo_{1_{ac}}&&&&&&Q_c
\end{diagram} 
 
\vspace{-4.14cm}\hspace{1.6cm}{\bf Lemma \ref{lm:compos}}

\vspace{-2.7cm}\hspace{6.6cm}{\bf Quantum\ teleportation} 
 
\vspace{7.1cm}

\newpage 

\noindent{\bf Proof of Theorem \ref{thm:logicgate} (logic-gate 
teleportation).} The top trapezoid
is the statement of  the Theorem. The $a$,
$b$ and $c$-labels are the same as above. For each $j\in\{1,2,3,4\}$ we have a diagram of the form
below. We ignore  the scalars -- which cancel out against each
other --  in this proof.
\vspace{0.5cm}
\begin{diagram}
\hspace{-1mm}Q&&&&&&\rTo^{\langle f\rangle_{i=1}^{i=4}\!\!\!}&&&&&&4\cdot Q\\
&\rdTo^{\,\rho_Q}&&&&&&&&&&\!\!\!\!\!\ruTo^{\bigoplus_{i=1}^{i=4}\varphi_i(f)^{-1}\!\!}&\\
&&\hspace{-2mm}Q\!\otimes\!{\rm I}&\rTo^{\!\!1_Q\!\otimes\!\uu
f\uuu\!}&\!Q\!\otimes\! Q^*\!\!\otimes\! Q\!&\rTo^{\!\langle\dd
\beta_i\ddd\rangle_{i=1}^{i=4}\!\otimes\!\!
1_Q\!\!}&(4\cdot {\rm
I})\!\otimes\!Q&\rTo^{\!\langle p_i^{\rm I}\!\otimes\!\!
1_Q\rangle_{i=1}^{i=4}}&
4\cdot \left({\rm I}\!\otimes\!
Q\right)&\rTo^{4\cdot \lambda^{-1}_Q\!}&4\cdot Q\hspace{-2mm}&&\\
\dIs&&&&&\hspace{-6.5mm}\rdTo_{\dd \beta_j\ddd\!\otimes\! 1_Q\hspace{-2mm}}&\dTo_{p^{\rm
I}_j\otimes\! 1_Q}&&\dTo_{p^{{\rm I}\otimes Q}_j}&&\dTo_{p_j^Q}&&\dTo_{p_j^Q}\\
\hspace{-1mm}Q&\rTo_{\!\!\!\!f\circ \beta_j}&Q&&\pile{\lTo^{\lambda^{-1}_Q}\\
\rTo_{\lambda_Q}}&&{\rm I}\otimes Q&\rTo_{1_{{\rm I}\otimes Q}}&{\rm
I}\otimes Q&\rTo_{\lambda^{-1}_Q}&Q&&\\ \uIs&\ruTo_{\!\!\varphi_j(f)\circ\!\! f\!\!}&&&&&&&&&&
\rdTo_{\varphi_j(f)^{-1}\!}\\
\hspace{-1mm}Q&&&&&&\rTo_{f}&&&&&&Q
\end{diagram} 
 
\vspace{-4.14cm}\hspace{1.6cm}{\bf Lemma \ref{lm:compos}}

\vspace{-2.7cm}\hspace{6.4cm}{\bf Logic-gate\ teleportation}

\vspace{7.1cm} 
\noindent{\bf Proof of Theorem \ref{thm:swap} (entanglement swapping).} The top trapezoid
is the statement of the Theorem. We have a diagram of the form below for each $j\in\{1,2,3,
4\}$. To simplify the notation of the types we set
$(a^*\!,b,c^*\!,d)$ for 
${Q_a^*\otimes Q_b\otimes Q_c^*\otimes Q_d}$ etc. We ignore the scalars  -- which cancel out 
against each other -- in this proof.
\vspace{0.5cm}
\begin{diagram}
{\rm I}\otimes{\rm I}&&&&&\rTo^{\langle\uu 1_{ba}\!\uuu\otimes\uu  
1_{dc}\!\uuu\rangle_{i=1}^{i=4}}&&&&&4\cdot (b^*\!,a,d^*\!,c)\\ 
&\rdTo^{\uu 1_{da}\!\uuu\!\otimes \!\uu 1_{bc}\!\uuu}&&&&&&&&\ruTo~{\sharp}&\\ 
&&(d^*\!,a,b^*\!,c)&\rTo^{\Theta_{ab}}&(d^*\!\!,4\!\cdot\!
(a,b^*),c)&\rTo^{(\tau,\upsilon)}&4\cdot(d^*\!,a,b^*\!,c)&\rTo^{4\cdot\sigma}& 
4\cdot(b^*\!,a,d^*\!,c)&&&\\ 
\uTo^{\rho_{\rm I}}&&&\rdTo~{\!1^*_d\otimes\dd\beta_j\ddd\otimes 1_c}&&\rdTo~{1_d^*\!\otimes p_j^{(a,b^*)}\!\otimes\!1_c}&\dTo_{p_j^{(d^*\!,a,b^*\!,c)}}& &\dTo_{p_j^{(b^*\!,a,d^*\!,c)}}&&\dTo_{p_j^{(b^*\!,a,d^*\!,c)}}\\ 
&&(d^*\!,c)&\lTo_{\rho^{-1}_{d^*}\!\otimes 1_c}&(d^*\!,{\rm
I}\,,c)&\rTo_{\hspace{-3mm}{\!1^*_d\!\otimes\!\uu\gamma_j\!\uuu\!\otimes\! 
1_c}\hspace{-3mm}}&(d^*\!,a,b^*\!,c)&\rTo_{\sigma}&(b^*\!,a,d^*\!,c)&&\\ 
&\ruTo^{\uu\beta_j\uuu\!\!\!}&\dTo_{\lambda_{d^*}\!\otimes
1_c}&\hspace{13mm}\ldTo_{\!\!\!\!\sigma^{-1}}&\dTo~{{\hspace{-5mm}1_{d}^*\!\otimes\!\uu 1_{ab\!}\!\uuu\!\otimes\!\beta_j^{-1}\hspace{-5mm}}}&&\dTo~{\hspace{-5mm}1_d^*\!\otimes\! \gamma_j^{\!-1}\!\!\otimes\!1_b^*\!\otimes\!\beta_j^{-1}}&&&\rdTo~{\ddagger}&\\ 
{\rm I}&&({\rm I}\,,d^*\!,c)&&(d^*\!,a,b^*\!,c)&\rIs&(d^*\!,a,b^*\!,c)&&\rTo_{\sigma}&&(b^*\!,a,d^*\!,c)\\
\dTo^{\lambda_{\rm I}}&&&\rdTo~{\uu 1_{ba\!}\!\uuu\!\otimes\!1_{d}^*\!\otimes\!\beta_j^{-1}}&\dTo_{\sigma^{-1}}\\
{\rm I}\otimes{\rm I}&&\rTo_{\uu 1_{ba}\!\uuu\otimes\uu 1_{dc}\!\uuu}&&(b^*\!,a,d^*\!,c)
\end{diagram} 
 
\vspace{-8.45cm}\hspace{6.4cm}{\bf Entanglement swapping} 

\vspace{4.44cm}\hspace{7.6cm}{\bf
Lemma \ref{lm:precompos}}

\vspace{1.44cm}\hspace{1.5cm}{\bf
Lemma \ref{lm:precompos}}

\vspace{-4.9cm}\hspace{1.0cm}{\bf  
Lemma \ref{lm:CUT}}

\vspace{4cm}\noindent
\hspace{9.6cm}$\sharp:= \bigoplus_{i=1}^{i=4}(1_b^*\!\otimes\!\gamma_i^{\!-1}\!\!\otimes\!\!1_d^*\!\otimes\!\beta_i^{-1}\!)$
\par\smallskip\noindent
\hspace{9.6cm}$\ddagger:=1_b^*\!\otimes\!\gamma_j^{\!-1}\!\!\otimes\!1_d^*\!\otimes\!\beta_j^{-1}$

\end{document}